\documentclass[showpacs,aps,prd,nofootinbib,floatfix,amsmath,amssymb, twocolumn]{revtex4}
\usepackage{graphicx}
\usepackage{hyperref}
\usepackage{float}
\restylefloat{table}

\newcommand{\al}			{\alpha}
\newcommand{\bt}			{\beta}
\newcommand{\gm}			{\gamma}
\newcommand{\dlt}			{\delta}

\newcommand{\tta}			{\theta}

\newcommand{\Dlt}			{\Delta}

\newcommand{\mc}[1]{\mathcal{ #1}}						
\newcommand{\trm}[1]{\textrm{ #1}}						
\newcommand{\mb}[1]{\mathbf #1}			  				
\newcommand{\ph}[1]{\phantom{ #1}}			  			

\newcommand{\wt}[1]{\widetilde{ #1}}						
\newcommand{\unl}[1]{\underline{ #1}}					

\newcommand{\ud}{\mathrm{d}} 								

\makeatletter
\newbox\slashbox \setbox\slashbox=\hbox{$/$}
\newbox\Slashbox \setbox\Slashbox=\hbox{\large$/$}
\def\pFMslash#1{\setbox\@tempboxa=\hbox{$#1$}
  \@tempdima=0.5\wd\slashbox \advance\@tempdima 0.5\wd\@tempboxa
  \copy\slashbox \kern-\@tempdima \box\@tempboxa}
\def\pFMSlash#1{\setbox\@tempboxa=\hbox{$#1$}
  \@tempdima=0.5\wd\Slashbox \advance\@tempdima 0.5\wd\@tempboxa
  \copy\Slashbox \kern-\@tempdima \box\@tempboxa}

\def\miss#1{\ifmmode{/\mkern-11mu #1}\else{${/\mkern-11mu #1}$}\fi}

\newcommand{\half}{\textstyle{\frac{1}{2}}}

\newcommand{\tbf}[1]{\textbf{#1}}														

\newcommand{\reff}[1]{~(\ref{#1})}															
\newcommand{\citte}[1]{~\cite{#1}}															

\newcommand{\cf}{\emph{cf.\,}}															
\newcommand{\eg}{\emph{e.g.\,}}														
\newtheorem{proposition}{Proposition}[section]	

\newcommand{\iA}[2]{{A}^{#1}_{#2}}

\newcommand{\psum}[1]{{\sum_{ #1}\!\!\!}'\,}

\numberwithin{equation}{section}				

\begin{document}

\makeatother

\title{Yang-Mills theories with an arbitrary number of compactified extra dimensions}

\author{ M. A. L\'{o}pez-Osorio$ {}^{(a)}$}\email{alopez@fcfm.buap.mx}
\author{E. Mart\'{i}nez-Pascual$ {}^{(a)}$}\email{emartinez@fcfm.buap.mx}
\author{H. Novales-S\' anchez${}^{(b)}$}\email{hnovales@fisica.ugto.mx}
\author{J. J. Toscano${}^{(a)}$}\email{jtoscano@fcfm.buap.mx}

\affiliation{$^{(a)}$Facultad de Ciencias F\'{\i}sico Matem\'aticas,
Benem\'erita Universidad Aut\'onoma de Puebla, Apartado Postal
1152, Puebla, Puebla, M\'exico.\\
$^{(b)}$Divisi\'{o}n de Ciencias e Ingenier\'{i}as, Universidad de Guanajuato Campus Le\'{o}n, Loma del Bosque 103, Colonia Lomas del Campestre, 37150, Le\'{o}n, Guanajuato, M\'{e}xico.
}

\begin{abstract}
The well-known Yang-Mills theory with one $ S^{1} / Z_{2}$ universal extra dimension (UED) is generalized to an arbitrary number of spatial extra dimensions through a novel compactification scheme. In this paper, the Riemannian flat base manifold under consideration contains $ n $ spatial extra dimensions defined by $ n $ copies of the orbifold $ S^{1} / Z_{2}$. In this approach, we present the gauge structure and the mass spectrum of the effective four dimensional theory. We introduce the concept of standard and nonstandard gauge transformations of the effective theory, and explicitly identify the emergence of massive vector fields in the same number as massless (`pseudo-Goldstone') scalars in the compactified theory, verifying that a Higgs-like mechanism operates in the compactification process. It is found that, in contrast with the one UED scenario, in cases with two or more UEDs there emerge massive scalar fields.  Besides, at a phase space level, the Hamiltonian analysis yields that the higher dimensional and compactified theories are classically equivalent using the fundamental concept of canonical transformation. This work lays the ground for a wider study on these theories concerning their quantization and predictive power at the level of quantum fluctuations.
\end{abstract}


\maketitle

\section{Introduction and background}
\label{in}
The concept of extra dimensions is a cornerstone of various theoretical frameworks that belong to the current \emph{Beyond Standard Model} research~\cite{L}.  Well motivated arguments~\cite{A,ADD,AADD} suggest that relatively large extra dimensions may become apparent at \emph{TeV}s scale, and hence effective field formulations become phenomenologically interesting. In the majority of scenarios with extra dimensions, our observed three dimensional space is regarded as a 3-brane embedded in a higher dimensional spacetime known as the bulk. The extra dimensional approach is defined depending on the bulk geometry and the way in which the different fields are allocated on it. In particular, allowing gauge and matter fields of the standard model to propagate in the bulk gives rise to  the universal extra dimensions (UEDs) scenarios~\cite{UED_first}; in this approach, gauge invariance preservation at the higher dimensional level requires all gauge parameters to  be also defined on the bulk~\citte{NT1}. In addition, to be consistent with the available experimental information, the submanifold corresponding to the extra dimensions is assumed to be suitably compactified.

An immediate effect of compactification is the emergence of Kaluza-Klein (KK) towers, that is, a variety of series into which fields and gauge parameters that propagate in the bulk expand. In each series, the zero modes of fields correspond to known fields of the standard model, whereas the zero modes of gauge parameters specify the so-called standard gauge transformations (SGTs)~\cite{NT1,LMNT}. These are gauge transformations underlying the lower dimensional theory, whose gauge parameters are the remainder of the gauge parameters defined on the bulk after compactification. In addition, each individual KK excitation of the fields becomes a well defined object with respect to the Poincar\'{e} group on four dimensions~\cite{LMNT}.

In recent years, the phenomenological implications of universal extra dimensions in relation to observables at a low-energy scale have been the subject of considerable interest in the context of dark matter~\cite{UEDDM}, neutrino physics~\cite{UEDn}, Higgs physics~\cite{UEDH}, flavor physics~\cite{UEDf}, hadronic and linear colliders~\cite{UEDc}, and electroweak gauge couplings~\cite{NT4}. Some theoretical aspects of these models have also been considered~\cite{EDT}. In this context, a comprehensive analysis of both the classical and quantum structure of Yang-Mills theories in the five dimensional $\left( S^{1}/Z_{2}\right) $-one UED (OUED) scenario was fully presented in a series of papers~\cite{NT1, LMNT, NT2}. 

In Ref.~\cite{NT1}, a new classification of the gauge invariance underlying the four dimensional compactified theory was introduced: The SGTs and the nonstandard gauge transformations (NSGTs). The latter were defined in terms of the KK excited modes of the gauge parameters that propagate in the bulk. The mass spectrum of the compactified theory --a Yang-Mills type theory manifestly invariant under $ SU(N,\mc{M}^{4}) $ and $ ISO(1,3) $-- presents one gauge vector field with components $ A^{(0)a}_{\mu} $, one KK tower of  massive vector fields with components $ A^{(k)a}_{\mu} $ ($ k>0 $), and one KK tower of massless (`pseudo-Goldtone') scalar fields with $ A^{(k)a}_{5} $ ($ k>0 $), where $ a $ stands for a gauge index.  The introduction of SGTs and NSGTs proved to be advantageous when quantizing the theory using Becchi-Rouet-Stora-Tyutin (BRST) techniques in its field-antifield formalism~\cite{BRST, FAF}. 

In Ref.~\cite{LMNT}, it was stressed that there exists a canonical transformation that maps a pure Yang-Mills theory defined on the $ m $ dimensional flat space $ \mc{M}^{m}=\mc{M}^{4}\times\mc{N}^{n} $
into the four dimensional theory resulting from compactification.
Therefore, compactification can be thought of as a mechanism that explicitly breaks down the $ ISO(1,m-1) $ symmetry into $ ISO(1,3) $, and simultaneously hides the $ SU(N,\mc{M}^{m}) $  gauge symmetry onto SGTs and NSGTs, the former being identified with $ SU(N,\mc{M}^{4}) $. Such a canonical transformation was constructed in detail within the OUED context, and it maps all relevant quantities defined in the higher dimensional theory (gauge symmetries, gauge fields, curvatures, Dirac constraints, the primary Hamiltonian and the gauge generator) onto the corresponding quantities in the four dimensional representation; this result implies the classical equivalence between the two theories. 

Although in the compactified theory there are pseudo-Goldstone bosons --thus allowing a Higgs-like mechanism to operate-- they do not correspond to genuine Goldstone bosons in the sense of the spontaneous breakdown of a global symmetry. The emergent pseudo-Goldstone bosons in Kaluza-Klein theories are directly generated by compactification of the extra spatial dimension. In the presence of the Higgs mechanism, a group $ G $ is broken down into a subgroup $ H $ and pseudo-Goldstone bosons become manifest as a consequence of the redefinition of fields about a chosen asymmetric vacuum. Both, this redefinition of fields and the compactification process can be thought as a part of canonical transformations when the theories are formulated in a Hamiltonian setting~\cite{LMNT}. It is worth noticing that in the spontaneous symmetry-breaking scenarios the remaining gauge subgroup $ H $ is generated by a proper subset of generators of $ G $, whereas in the compactification process, the remaining manifest gauge-invariance group $ SU(N,\mc{M}^{4}) $ arises from $ SU(N,\mc{M}^{m}) $ once one integrates out the extra dimensions in the gauge parameters related to the higher-dimensional theory.

To our knowledge, in the literature there has been plenty of research devoted to models with the minimal integer number of UEDs in contrast with higher number of universal extra-dimensional models. In fact, it is known that slight deviations above the five dimensional one-UED context cause important differences in the phenomenology of the models themselves; for an up-to-date review see Ref.~\cite{Se} and references therein. In the present work, those results reported in Ref.~\cite{LMNT} will be generalized to an arbitrary number of extra dimensions using a novel  generalization of models with $ S^{1}/Z_{2} $ as universal spatial extra dimension. In this paper the base manifold under consideration contains $ n $ extra spatial dimensions defined on  $ \underbrace{S^{1}/Z_{2}\times S^{1}/Z_{2}\times\ldots\times S^{1}/Z_{2}}_{n\,\trm{times}} $. Using Yang-Mills theories, this assumption lets us explicitly write down the canonical transformation that maps all relevant fields in the higher-dimensional theory onto the fields  that naturally describe the lower-dimensional effective theory. By construction, the higher dimensional theory contains fields with well-defined transformation laws under the groups  $ SU(N,\mc{M}^{m}) $ and $ ISO(1,m-1) $, and the fields in the lower-dimensional theory will correspond to objects with well-defined transformation laws under the groups $ SU(N,\mc{M}^{4}) $ and $ ISO(1,3) $. In this setting, it will also be shown that the gauge structure of the higher-dimensional theory is duly mapped onto that of the effective theory formed by SGTs and NSGTs; this is done by showing that the canonical transformation permeates at all levels in the Dirac-Bergmann algorithm for constraints~\cite{dirac_etal}. This result implies  the classical equivalence between the effective and the original theory. Using this compactification scheme, a precise study of the mass spectrum in the compactified theory can be achieved. In general, the masses of the KK modes receive contributions from bulk radiative corrections and boundary terms~\cite{CMS}. These effects yield a splitting of the near degeneracy of KK modes at tree level and violate KK number. It has been shown~\cite{CMS2} that such contributions, which depend on the cutoff scale of the extra-dimensional description, may have important phenomenological implications. We would like to point out that in the present paper we focus on the masses of the KK modes at tree level, so whenever we say {\it mass spectrum}, we actually refer to {\it tree-level mass spectrum}. It will be seen that after various unitary transformations in the space of basic fields, the compactified Yang-Mills theory --manifestly invariant under the gauge group $ SU(N,\mc{M}^4) $ and $ ISO(1,3) $-- contains one gauge vector field, $ (2^{n}-1) $ KK towers of massive vector fields, $ (2^{n}-1) $ KK towers of massless (`pseudo-Goldstone') scalar fields, and $ (2^{n}-1)(n-1) $ massive scalar fields. Contrasting the number of KK towers of massive $ ISO(1,3) $ vector fields and $ ISO(1,3) $ massless scalar fields, we conclude that a Higgs-like mechanism also operates in the compactification of $ n $ extra spatial dimensions in pure Yang-Mills theories. 

As it was emphasized in Refs.~\cite{NT1,NT2,NT4}, these theories must be consistently quantized since the excited KK states contribute to standard Green's functions (SGF) at the one--loop level, that is, to Green's functions with only zero modes in external legs. A remarkable result reported in these references was the proof that contributions from the extra dimension to the SGF are renormalizable at one--loop level. The only type of divergences are those already present in the standard theory (the theory with only zero modes), and hence they can be absorbed in the ordinary way~\cite{NT1,NT2}. All these considerations translate automatically to the standard model in the OUED scenario~\cite{NT4,CGBT}. From the theoretical viewpoint, it is interesting to investigate if similar results can be obtained in the quantization of the model with an arbitrary number of extra dimensions. Besides, according to the phenomenological understanding, it is relevant to explore the one-loop renormalizability of the SGF in this scenario and eventually analyse the sensitivity to the number of UEDs of various standard model observables at low-energy scales in the compactification scheme that is proposed in this paper. In fact, these are the final goals of our investigation. In this paper we will focus on the classical structure of the compactified gauge theory, its quantization and one-loop renormalizability will be tackled in a next communication~\cite{LMNT2}. In this way, all concepts introduced in Ref.~\cite{LMNT} and in the present paper will be lifted to a practical level to search physics beyond the standard model.

The rest of the paper has been organized as follows. Sec.~\ref{g}. contains most of the new results presented in this paper. In section \ref{ss:A}, the following is included: the compactification scheme for the $ m $-dimensional pure $ SU(N,\mc{M}^{m}) $ Yang-Mills theory, the Lagrangian structure of the compactified theory and the classification of gauge invariances of the theory into SGTs and NSGTs, and a  detailed analysis of the mass spectrum of the effective theory. In section \ref{ss:B}, the canonical transformation that connects the higher- and four-dimensional theories is presented; moreover, it also contains the gauge structure analysis of the compactified theory in phase space using the Dirac-Bergmann algorithm for singular systems, and the gauge generator of the SGTs and NSGTs are separately presented. Due to the convoluted notation necessary to perform the generalization to an arbitrary number of dimensions, Sec.~\ref{examples} is devoted to the specialization of the general case to the $ m=5$, $m=6 $ and $ m=7 $ cases. Section \ref{conclu} presents a summary and concluding remarks. Finally, in the Appendix \ref{App1}, various tables that summarize the mass spectrum of the particular cases $ m=6 $ and $ m=7 $ are collected.
\vspace*{-.2cm}

\section{Yang-Mills theories with an arbitrary number of UED\texorpdfstring{\lowercase{s}}{}}
\label{g}

In this section the gauge structure analysis of the Yang-Mills theory with one UED as presented in Refs.~\cite{NT1,LMNT} is extended to a theory with an arbitrary number of UEDs. We will focus our attention on various classical aspects of the compactified action. First of all, we will study the gauge transformations of basic fields at the configuration-space level and classify them into SGTs and NSGTs. These transformations are central to quantizing gauge theories in the Lorentz-covariant context of the field-antifield formalism of BRST~\cite{NT1,BRST}.  Secondly, we will show a detailed analysis of the mass spectrum presented in the compactified theory, from where we will conclude that a Higgs mechanism operates as a consequence of the compactification. Thirdly, the Hamiltonian analysis will be carried out, yielding the gauge generators of the theory as well as the proof that the ``fundamental'' and compactified theories are classically equivalent.

\subsection{Lagrangian structure}
\label{ss:A}
The higher-dimensional theory to compactify is a pure Yang-Mills theory on a flat Riemannian manifold $ \mc{M}^{m}= \mc{M}^{4}\times\mc{N}^{n} $ based on the $ SU(N) $ gauge group, which will be denoted by $ SU(N,\mc{M}^{m}) $ in order to emphasize that gauge parameters $ \al^{a} $ are allowed to be defined over all $ \mc{M}^{m} $.  The components of the connection $ \mc{A} $ defined on $ \mc{M}^{m} $, also referred to as gauge fields, will be denoted by $ \mc{A}^{a}_{M}(x,y) $ once evaluated at each point $ (x,y) $ of $ \mc{M} ^{m}$. The gauge index $ a $ runs from 1 to $ N^{2}-1 $. The pure Yang-Mills action on the manifold $  \mc{M}^{m} $ is constructed through the Lagrangian
\begin{equation}\label{YML}
\mathcal{L}_{\!\trm{YM}}(x,y)=-\frac{1}{4}\mathcal{F}_{MN}^{a}(x,y)\mathcal{F}_{a}^{MN}(x,y)\ ,
\end{equation}
where repeated indices imply summation, and $ M,N=0,1,2,3,5,\ldots,m $. We wish to recover the five dimensional case reported in Refs.~\cite{NT1,LMNT} by setting $ m=5 $, or equivalently, $ n=1 $. The Lagrangian \eqref{YML} is invariant under the following gauge transformations
\begin{equation}\label{gt-n}
\delta\mc{A}^{a}_{M}=\mc{D}^{ab}_{M}\al ^{b}\ ,
\end{equation}
where $\mc{D}^{ab}_{M}=\dlt^{ab}\partial_{M}-g_{m}f^{abc}\mc{A}^{c}_{M}$ is the covariant derivative in the adjoint representation, $\al^{a}$ are infinitesimal gauge parameters on $ \mc{M}^{m} $, and the coupling constant $ g_{m} $ has dimensions of (mass)$^{(4-m)/2} $. The curvature components $ \mathcal{F}_{MN}^{a} $ in terms of gauge fields are given by
\begin{align}\label{eq:FMNa}
\mathcal{F}^{a}_{MN}(x,y)=& \ \partial_{M}\mathcal{A}^{a}_{N}(x,\,y)-\partial_{N}\mathcal{A}_{M}^{a}(x,\,y)\nonumber\\
& \ +g_{m}f_{abc}\,\mathcal{A}_{M}^{b}(x,\,y)\mathcal{A}_{N}^{c}(x,\,y)\ ,
\end{align}
and the gauge transformations Eq.~\eqref{gt-n} imply  the following variations on the curvature components:
\begin{equation}\label{gtF-n}
\dlt\mc{F}^{a}_{MN}(x,y)=g_{m}f^{abc}\mc{F}^{b}_{MN}(x,y)\al^{c}(x,y)\ .
\end{equation}
According to their transformation law with respect to the Lorentz group $SO(1,3)$, the gauge fields $\mathcal{A}_{M}^{a}(x,\,y)$ split into 4-vectors, $\mathcal{A}_{\mu}^{a}(x,\,y)$, and scalars $\mathcal{A}_{\bar{\mu}}^{a}(x,\,y)$ (see Ref.~\cite{LMNT}). From now on, $ \bar{\mu}=5,6,\ldots, m$ will enumerate extra dimensions; in this notation, $(x,y)=(x^{0},x^{1},x^{2},x^{3},$ $y^{1},y^{2},\ldots,y^{n}) =(x^{0},x^{1},x^{2},x^{3},x^{5},x^{6},\ldots,x^{m})=(x^{\mu},x^{\bar{\mu}}) $ are coordinates of $ \mc{M}^{m} =\mc{M}^{4}\times \mc{N}^{n}$.

We regard $ \mc{N}^{n} $ as the product of $ n $ circles of different radii $ R_{1},\ldots,R_{n} $, and assume the following periodicity conditions on gauge fields and gauge parameters
\begin{subequations}\label{per-cond-n}
\begin{align}
\mathcal{A}_{M}^{a}(x,y+R)=\mathcal{A}_{M}^{a}(x,y) \ , \label{A-periodic-n}\\
\al^{a}(x,y+R)=\al^{a}(x,y)\ , \label{al-periodic-n}
\end{align}
\end{subequations}
where $ R=(R_{1},\ldots,R_{n}) $. Replacing each extra dimension with the corresponding circle by the action of $Z_{2}$ so that, $ y $ is identified with its antipode on $ S^{1} $, a pure four-dimensional Yang-Mills sector within the effective theory is recovered after compactification. Henceforth, we assume the following parity conditions on gauge fields and gauge parameters:
\begin{subequations}\label{par-cond-n}
\begin{align}
\mathcal{A}_{\mu}^{a}(x,-y)&=\mathcal{A}_{\mu}^{a}(x,y)\ , \label{paAv}\\
\mathcal{A}_{\bar{\mu}}^{a}(x,-y)&=-\mathcal{A}_{\bar{\mu}}^{a}(x,y)\ , \label{paAs}\\
\al^{a}(x,-y)&=\al^{a}(x,y)\ .  \label{al-par-n}
\end{align}
\end{subequations}

\begin{widetext}
As in the framework developed in Ref.~\cite{LMNT}, the set of requirements (\ref{A-periodic-n}), (\ref{paAv}, and~\ref{paAs}) permits us to write a point transformation which maps $SU(N,{\cal M}^m)$  gauge fields $ \mc{A}^{a}_{M} $ onto $SU(N,{\cal M}^4)$ gauge fields and other fields with well-defined transformation laws under $SU(N,{\cal M}^4)$ and $ ISO(1,3) $. The group $ SU(N,\mc{M}^{4}) $ can be regarded as a subgroup of  $SU(N,{\cal M}^m)$ as it can be explicitly obtained by an appropriate reduction of  the gauge parameters $ \al^{a}$ to the submanifold $ \mc{M}^{4} $. The aforementioned transformation corresponds to the following Fourier expansions:
\begin{subequations}\label{AM-expand}
\begin{align}
\mc{A}^{a}_{\mu}(x,y)  =&\ \left(1/\prod_{\al=1}^{n}R_{\al}\right)^{1/2}{A}^{(0,\ldots, 0)a}_{\mu}(x) \nonumber\\
&+ \left(2/\prod_{\al=1}^{n}R_{\al}\right)^{1/2}{\sum_{\unl{m}_{1}\ldots \unl{m}_{n}}\!\!\!}'\,{A}^{(\unl{m}_{1},\ldots , \unl{m}_{n})a}_{\mu}(x)\cos\left[ 2\pi\left(\frac{\unl{m}_{1}y^{1}}{R_{1}}+\cdots+\dfrac{\unl{m}_{n}y^{n}}{R_{n}}\right)\right]\ ,\label{Am-expand}\\
\mc{A}^{a}_{\bar{\mu}}(x,y) =&\  \left(2/\prod_{\al=1}^{n} R_{\al}\right)^{1/2} {\psum{\unl{m}_{1}\ldots \unl{m}_{n}}} {A}^{(\unl{m}_{1},\ldots , \unl{m}_{n})a}_{\bar{\mu}}(x)\sin\left[2\pi\left(\frac{\unl{m}_{1}y^{1}}{R_{1}}+\cdots+\dfrac{\unl{m}_{n}y^{n}}{R_{n}}\right)\right]\ .\label{Abm-expand}
\end{align}
\end{subequations}
Some remarks on the notation are in order here. The primed sum $ {\sum}_{\unl{m}_{1},\ldots,\unl{m}_{n}} ' $ takes into account all the Fourier-mode combinations except the zero mode $ (0,0,\ldots,0) $; for example, in the $ n=3 $ case, given two quantities $S^{(\unl{m}_{1},\unl{m}_{2},\unl{m}_{3})}  $ and $ T^{(\unl{m}_{1},\unl{m}_{2},\unl{m}_{3})} $, the primed sum can be arranged into 7 different terms which are sums over repeated Fourier indices. Explicitly
\begin{align}\label{eq:primedsum}
 \psum{\unl{m}_{1}\unl{m}_{2}\unl{m}_{3}}S^{(\unl{m}_{1},\unl{m}_{2},\unl{m}_{3})}\,T^{(\unl{m}_{1},\unl{m}_{2},\unl{m}_{3})} & =\sum_{m_{1}=1}^{\infty}S^{(m_{1},0,0)}\,T^{(m_{1},0,0)}+\sum_{m_{2}=1}^{\infty}S^{(0,m_{2},0)}\,T^{(0,m_{2},0)}
 +\sum_{m_{3}=1}^{\infty}S^{(0,0,m_{3})}\,T^{(0,0,m_{3})}\nonumber\\
 &  +\sum_{m_{1},m_{2}=1}^{\infty} S^{(m_{1},m_{2},0)}\,T^{(m_{1},m_{2},0)}+\sum_{m_{1},m_{3}=1}^{\infty} S^{(m_{1},0,m_{3})}\,T^{(m_{1},0,m_{3})}\nonumber\\
 &  +\sum_{m_{2},m_{3}=1}^{\infty} S^{(0,m_{2},m_{3})}\,T^{(0,m_{2},m_{3})}+\sum_{m_{1},m_{2,}m_{3}=1}^{\infty} S^{(m_{1},m_{2},m_{3})}\,T^{(m_{1},m_{2},m_{3})}\ .
\end{align}
\end{widetext}
The primed sum over $ n $ Fourier indices can be arranged into  $ 2^{n}-1 $ different terms which are conventional sums, all of them starting from 1. The zero Fourier mode ${A}^{(0,\ldots, 0)a}_{\mu}$ in the expansion \eqref{Am-expand}  corresponds to the components of  a $SU(N,{\cal M}^4)$ gauge vector field, whereas all excited KK modes ${A}^{(m_{1},0,\ldots ,0)a}_{M}$, ${A}^{(0,m_{2},0,\ldots , 0)a}_{M}$, $ \ldots $ , ${A}^{(0,0,\ldots , m_{n})a}_{M}$, ${A}^{(m_{1},m_{2},0,\ldots , 0)a}_{M}$, ${A}^{(m_{1},0,m_{3},0,\ldots , 0)a}_{M}$, $ \ldots $ , ${A}^{(m_{1},\ldots , m_{n})a}_{M}$ transform under the adjoint representation of this group [see \eqref{AnSGT} below].

Taking advantage of the periodicity and parity properties \eqref{al-periodic-n} and \eqref{al-par-n} of the gauge parameters, and using the primed sum notation, the corresponding Fourier expansions for these functions take the form
\begin{align}
\alpha^{a}(x,y) =&\ \left(1/\prod_{\al=1}^{n}R_{\al}\right)^{1/2}\alpha^{(0,\ldots, 0)a}(x) \nonumber\\
&+ \left(2/\prod_{\al=1}^{n}R_{\al}\right)^{1/2}{\sum_{\unl{m}_{1}\ldots \unl{m}_{n}}\!\!\!}'\,\alpha^{(\unl{m}_{1},\ldots , \unl{m}_{n})a}(x)\nonumber\\
& \times\cos\left[ 2\pi\left(\frac{\unl{m}_{1}y^{1}}{R_{1}}+\cdots+\dfrac{\unl{m}_{n}y^{n}}{R_{n}}\right)\right]\label{Fsgp}\ .
\end{align}
Notice that if the dependence of gauge parameters is reduced to $ \mc{M}^{4}$ by integrating out the extra spatial dimensions, then only the  scalar functions $\alpha^{(0,\ldots, 0)a}$ remain in this expansion. It is worth to mention  that in the BRST formalism,  gauge parameters are elevated to degrees of freedom at the classical level since they are needed to quantize the theory.

On the spacetime manifold $ \mc{M}^{4}\times\mc{N}^{n} $, the curvature components belong to three qualitatively different sectors with respect to their transformation law under $ ISO(1,3) $: (i) the $ (\mc{M}^{4}$-$\mc{M}^{4} )$ sector, $ \mathcal{F}^{a}_{\mu\nu}(x,y) $, contains 2-tensors; (ii) the $ (\mc{M}^{4}$-$\mc{N}^{n} )$ sector, $ \mathcal{F}^{a}_{\mu\bar{\nu}}(x,y)  $, contains vectors; and (iii) the $ (\mc{N}^{n}$-$\mc{N}^{n})$ sector, $ \mathcal{F}^{a}_{\bar{\mu}\bar{\nu}}(x,y) $, contains scalars. The periodicity and parity properties of the curvature components, after compactification, are dictated by the corresponding properties of the gauge fields; therefore, 
\begin{equation}
\mc{F}^{a}_{MN}(x,y+R)=\mc{F}^{a}_{MN}(x,y)\ , \label{F-periodic-n}
\end{equation}
 and
\begin{subequations}\label{FMN-par-n}
\begin{align}
\mc{F}^{a}_{\mu\nu}(x,-y) & =  \mc{F}^{a}_{\mu\nu}(x,y)\ ,\label{eq:Fparity1}\\
\mc{F}^{a}_{\mu\bar{\nu}}(x,-y) & = -\mc{F}^{a}_{\mu\bar{\nu}}(x,y)\ ,\label{eq:Fparity2}\\
\mc{F}^{a}_{\bar{\mu}\bar{\nu}}(x,-y) & = \mc{F}^{a}_{\bar{\mu}\bar{\nu}}(x,y)\ .\label{eq:Fparity3}
\end{align}
\end{subequations}
 \begin{widetext}
The set of requirements\reff{F-periodic-n} and  \eqref{FMN-par-n} allows us to expand the different curvature sectors in Fourier series as follows:
\begin{subequations}\label{FMN-expand-n}
\begin{align}
\mc{F}^{a}_{\mu\nu}(x,y)  =&\ \left(1/\prod_{\al=1}^{n}R_{\al}\right)^{1/2}\mc{F}^{(0,\ldots, 0)a}_{\mu\nu}(x) \nonumber\\
&+ \left(2/\prod_{\al=1}^{n}R_{\al}\right)^{1/2}{\psum{\unl{m}_{1}\ldots \unl{m}_{n}}}\mc{F}^{(\unl{m}_{1},\ldots , \unl{m}_{n})a}_{\mu\nu}(x)\cos\left[ 2\pi\left(\frac{\unl{m}_{1}y^{1}}{R_{1}}+\cdots+\dfrac{\unl{m}_{n}y^{n}}{R_{n}}\right)\right]\ ,\label{Fmn-expand-n}\\
\mc{F}^{a}_{\mu\bar{\nu}}(x,y)  =&\  \left(2/\prod_{\al=1}^{n} R_{\al}\right)^{1/2} {\psum{\unl{m}_{1}\ldots \unl{m}_{n}}} \mc{F}^{(\unl{m}_{1},\ldots , \unl{m}_{n})a}_{\mu\bar{\nu}}(x)\sin\left[2\pi\left(\frac{\unl{m}_{1}y^{1}}{R_{1}}+\cdots+\dfrac{\unl{m}_{n}y^{n}}{R_{n}}\right)\right]\ , \label{Fmbm-expand-n}\\
\mc{F}^{a}_{\bar{\mu}\bar{\nu}}(x,y)  =&\ \left(1/\prod_{\al=1}^{n}R_{\al}\right)^{1/2}\mc{F}^{(0,\ldots, 0)a}_{\bar{\mu}\bar{\nu}}(x) \nonumber\\
&+ \left(2/\prod_{\al=1}^{n}R_{\al}\right)^{1/2}{\psum{\unl{m}_{1}\ldots \unl{m}_{n}}}\mc{F}^{(\unl{m}_{1},\ldots , \unl{m}_{n})a}_{\bar{\mu}\bar{\nu}}(x)\cos\left[ 2\pi\left(\frac{\unl{m}_{1}y^{1}}{R_{1}}+\cdots+\dfrac{\unl{m}_{n}y^{n}}{R_{n}}\right)\right] \ .\label{Fbmbn-expand-n}
\end{align}
\end{subequations}
\end{widetext}

The orthogonality of the trigonometric functions enables a direct integration of the compact extra dimensions in Eq.~\eqref{YML}, obtaining in this way an effective four-dimensional theory
\begin{equation}
\mc{L}_{\!\trm{4YM}}=\int_{0}^{R_{1}}\cdots\int_{0}^{R_{n}}\ud^{n}y\ \mc{L}_{\!\trm{YM}}\ .
\end{equation}
The effective Lagrangian becomes a function of the zero modes $ \mc{F}^{(0,\ldots, 0)a}_{\mu\nu} $ and $ \mc{F}^{(0,\ldots, 0)a}_{\bar{\mu}\bar{\nu}} $, as well as all the excited modes  $ \mc{F}^{(0,m_{i},0,\ldots,0)a}_{MN} $, $ \mc{F}^{(0,m_{i},0,\ldots,0,m_{j},0)a}_{MN} $, $ \ldots $ , $ \mc{F}^{(m_{1},m_{2},\ldots,m_{n})a}_{MN} $. Explicitly
\begin{align}\label{eq:L4YM}
\mc{L}_{\!\trm{4YM}} = &\ -\dfrac{1}{4}\Big[\mc{F}^{(0,\ldots,0)a}_{\mu\nu}\mc{F}^{(0,\ldots,0)\mu\nu}_{a}+\mc{F}^{(0,\ldots,0)a}_{\bar{\mu}\bar{\nu}}\mc{F}^{(0,\ldots,0)\bar{\mu}\bar{\nu}}_{a}\big.\Big.\nonumber\\
&\ +{\psum{\unl{m}_{1}\ldots \unl{m}_{n}}}\big(\mc{F}^{(\unl{m}_{1},\ldots,\unl{m}_{n})a}_{\mu\nu}\mc{F}^{(\unl{m}_{1},\ldots,\unl{m}_{n})\mu\nu}_{a}
\nonumber\\
&  \Big.\big.\ +2\,\mc{F}^{(\unl{m}_{1},\ldots,\unl{m}_{n})a}_{\mu\bar{\nu}}\mc{F}^{(\unl{m}_{1},\ldots,\unl{m}_{n})\mu\bar{\nu}}_{a}\nonumber\\
&\ +\mc{F}^{(\unl{m}_{1},\ldots,\unl{m}_{n})a}_{\bar{\mu}\bar{\nu}}\mc{F}^{(\unl{m}_{1},\ldots,\unl{m}_{n})\bar{\mu}\bar{\nu}}_{a}\, \big)\Big]\ ,
\end{align}
notice that, after expanding the primed sum, no coupling between curvature components with two different Fourier modes occurs in the Lagrangian.

By virtue of Eqs.~\eqref{eq:FMNa},~\eqref{AM-expand} and~\eqref{FMN-expand-n}, the effective action $ S=\int\ud^{4}x \,\mc{L}_{\!\trm{4YM}}$ becomes a functional of the zero mode $ {A}^{(0,\ldots,0)a}_{\mu} $ and all the excited KK modes ${A}^{(0,m_{i},0,\ldots , 0)a}_{M}$, $ \ldots $ , $ {A}^{(m_{1},\ldots,m_{n})a}_{M} $. These are the built-in configuration variables of the compactified theory. Indeed, it is straightforward to determine the form of $ \mc{F}^{(0,\ldots, 0)a}_{\mu\nu} $, $ \mc{F}^{(0,\ldots, 0)a}_{\bar{\mu}\bar{\nu}} $, and all excited KK modes $ \mc{F}^{(0,m_{i},0,\ldots,0)a}_{MN} $, $ \ldots$ , $ \mc{F}^{(m_{1},\ldots,m_{n})a}_{MN} $ in terms of $ {A}^{(0,\ldots,0)a}_{\mu} $ and all excited KK modes of gauge fields $ {A}^{(0,m_{i},0,\ldots,0)a}_{M} $, $ \ldots $ , ${A}^{(m_{1},\ldots , m_{n})a}_{M}$: One introduces the expansions~\eqref{AM-expand} into the explicit form of the curvature as a function of gauge fields\reff{eq:FMNa}, then uses the Fourier expansions of the curvature sectors  \eqref{FMN-expand-n}. Then --by taking advantage of the orthogonality of the trigonometric functions-- one obtains (after convenient integrations)
\begin{subequations}\label{eq:FofA-fourier}
\begin{align}
\mc{F}^{(0,\ldots, 0)a}_{\mu\nu} = &\  F^{(0,\ldots,0)a}_{\mu\nu}+gf_{abc}\nonumber\\
& \ \times\!\!{\psum{\, \,\unl{k}_{1},\ldots \unl{k}_{n}\,\,}}\iA{(\unl{k}_{1},\ldots,\unl{k}_{n})b}{\mu}\iA{(\unl{k}_{1},\ldots,\unl{k}_{n})c}{\nu} \ , \label{eq:FofA-fourier1}\\
\mc{F}^{(0,\ldots, 0)a}_{\bar{\mu}\bar{\nu}} = &\  gf_{abc}\!\!{\psum{\, \,\unl{k}_{1}\ldots \unl{k}_{n}\,\,}}\iA{(\unl{k}_{1},\ldots,\unl{k}_{n})b}{\bar{\mu}}\iA{(\unl{k}_{1},\ldots,\unl{k}_{n})c}{\bar{\nu}} \ , \label{eq:FofA-fourier2}
\end{align}
\end{subequations}
and all the excited KK modes summarized as follows:
\begin{widetext}
\begin{subequations}\label{FKKmodes}
\begin{align}
\mc{F}^{(\unl{m}_{1},\ldots,\unl{m}_{n})a}_{\mu\nu}   = &\ 2\mc{D}^{(0,\ldots,0)ab}_{[\mu}\iA{(\unl{m}_{1},\ldots,\unl{m}_{n})b}{\nu]}\nonumber\\
&\ + gf_{abc}{\sum_{\unl{r}_{1}\ldots \unl{r}_{n}}\!}'\ {\sum_{\unl{k}_{1}\ldots \unl{k}_{n}}\!}' \iA{(\unl{k}_{1},\ldots,\unl{k}_{n})b}{\mu}\iA{(\unl{r}_{1},\ldots,\unl{r}_{n})c}{\nu}{\Dlt}_{\unl{m}_{1}\ldots \unl{m}_{n}\unl{k}_{1}\ldots \unl{k}_{n}\unl{r}_{1}\ldots \unl{r}_{n}}\ ,\label{eq:FofA-fourier3}\\
\mc{F}^{(\unl{m}_{1},\ldots,\unl{m}_{n})a}_{\mu\bar{\nu}}  = &\ \mc{D}^{(0,\ldots,0)ab}_{\mu}\iA{(\unl{m}_{1},\ldots,\unl{m}_{n})b}{\bar{\nu}} +2\pi\left(\dfrac{
\unl{m}_{1}\dlt_{\bar{\nu}5}}{R_{1}}+\cdots+\dfrac{\unl{m}_{n}\dlt_{\bar{\nu}\, n+4}}{R_{n}}\right)\iA{(\unl{m}_{1},\ldots,\unl{m}_{n})a}{{\mu}}\nonumber\\
& + gf_{abc}{\sum_{\unl{r}_{1}\ldots \unl{r}_{n}}\!}'\ {\sum_{\unl{k}_{1}\ldots \unl{k}_{n}}\!}' \iA{(\unl{k}_{1},\ldots,\unl{k}_{n})b}{\mu}\iA{(\unl{r}_{1},\ldots,\unl{r}_{n})c}{\bar{\nu}}{\Dlt}'_{\unl{m}_{1}\ldots \unl{m}_{n}\unl{r}_{1}\ldots \unl{r}_{n}\unl{k}_{1}\ldots \unl{k}_{n}}\ ,\label{eq:FofA-fourier4}\\
\mc{F}^{(\unl{m}_{1},\ldots,\unl{m}_{n})a}_{\bar{\mu}\bar{\nu}} = & \ 4\pi\left(\dfrac{
\unl{m}_{1}\dlt_{5[\bar{\mu}}}{R_{1}}+\cdots+\dfrac{\unl{m}_{n}\dlt_{n+4\, [\bar{\mu}}}{R_{n}}\right)\iA{(\unl{m}_{1},\ldots,\unl{m}_{n})a}{{\bar{\nu}]}}\nonumber\\
& \ + gf_{abc}{\sum_{\unl{r}_{1}\ldots \unl{r}_{n}}\!}'\ {\sum_{\unl{k}_{1}\ldots \unl{k}_{n}}\!}' \iA{(\unl{k}_{1},\ldots,\unl{k}_{n})b}{\bar{\mu}}\iA{(\unl{r}_{1},\ldots,\unl{r}_{n})c}{\bar{\nu}}{\Dlt}'_{\unl{k}_{1}\ldots \unl{k}_{n}\unl{r}_{1}\ldots \unl{r}_{n}\unl{m}_{1}\ldots \unl{m}_{n}}\ .\label{eq:FofA-fourier5}
\end{align}
\end{subequations}
From now on, equations for Fourier components labeled by $ (\unl{m}_{1},\unl{m}_{2},\ldots,\unl{m}_{n}) $ are valid for any combination of non-negative integers except $ (0,0,\ldots,0) $. Hence, for example, in Eq.~\eqref{eq:FofA-fourier4} the functional form of $ \mc{F}^{(0,m_2,0,\ldots,0)a}_{\mu\bar{\nu}} $ in terms of gauge-field components can be found. Moreover, enclosing two indices within a squared bracket implies antisymmetrization, \eg ~$ T_{[\mu\nu]}:=\half (T_{\mu\nu}-T_{\nu\mu}) $, and the following definitions are used:
\begin{subequations}\label{eq:deltas}
\begin{align}
{\Dlt}_{\unl{k}_{1}\ldots \unl{k}_{n}\unl{r}_{1}\ldots \unl{r}_{n}\unl{m}_{1}\ldots \unl{m}_{n}}\equiv &\dfrac{1}{\sqrt{2}}\left(\dlt_{\unl{k}_{1}\,\unl{r}_{1}+\unl{m}_{1}}\cdots \dlt_{\unl{k}_{n}\,\unl{r}_{n}+\unl{m}_{n}}+\dlt_{\unl{r}_{1}\,\unl{k}_{1}+\unl{m}_{1}}\cdots\dlt_{\unl{r}_{n}\,\unl{k}_{n}+\unl{m}_{n}}+\dlt_{\unl{m}_{1}\,\unl{k}_{1}+\unl{r}_{1}}\cdots\dlt_{\unl{m}_{n}\,\unl{k}_{n}+\unl{r}_{n}}\right)\ , \label{eq:delta}\\
{\Dlt}'_{\unl{k}_{1}\ldots \unl{k}_{n}\unl{r}_{1}\ldots \unl{r}_{n}\unl{m}_{1}\ldots \unl{m}_{n}}\equiv & \dfrac{1}{\sqrt{2}}\left(\dlt_{\unl{k}_{1}\,\unl{r}_{1}+\unl{m}_{1}}\cdots \dlt_{\unl{k}_{n}\,\unl{r}_{n}+\unl{m}_{n}}+\dlt_{\unl{r}_{1}\,\unl{k}_{1}+\unl{m}_{1}}\cdots\dlt_{\unl{r}_{n}\,\unl{k}_{n}+\unl{m}_{n}}-\dlt_{\unl{m}_{1}\,\unl{k}_{1}+\unl{r}_{1}}\cdots\dlt_{\unl{m}_{n}\,\unl{k}_{n}+\unl{r}_{n}}\right)\ . \label{eq:delta'}
\end{align}
\end{subequations}
\end{widetext}
Although these multi-index objects are well defined for all positive values of their indices, including the case where all of them vanish, the values $ \Dlt_{0\ldots 0} $ and $ \Dlt'_{0\ldots 0} $ are not needed since they do not arise in Eqs.~\eqref{FKKmodes}. As a consequence of their definition,  the multi-index objects $ {\Dlt} $ and $ {\Dlt}' $ have the following symmetry properties in their indices:
\begin{subequations}\label{dlts-props}
\begin{align}
{\Dlt}_{\unl{k}_{1}\ldots \unl{k}_{n}\unl{r}_{1}\ldots \unl{r}_{n}\unl{m}_{1}\ldots \unl{m}_{n}} & = {\Dlt}_{\unl{m}_{1}\ldots \unl{m}_{n}\unl{k}_{1}\ldots \unl{k}_{n}\unl{r}_{1},\ldots \unl{r}_{n}}\nonumber\\
&={\Dlt}_{\unl{r}_{1},\ldots \unl{r}_{n}\unl{m}_{1}\ldots \unl{m}_{n}\unl{k}_{1}\ldots \unl{k}_{n}}\ , \label{dlts-props1}\\
{\Dlt}'_{\unl{k}_{1}\ldots \unl{k}_{n}\unl{r}_{1}\ldots \unl{r}_{n}\unl{m}_{1}\ldots \unl{m}_{n}} & ={\Dlt}'_{\unl{r}_{1}\ldots \unl{r}_{n}\unl{k}_{1}\ldots \unl{k}_{n}\unl{m}_{1}\ldots \unl{m}_{n}} \ . \label{dlts-props2}
\end{align}
\end{subequations}
The following observation is useful in the calculation of some values taken by these multi-index object. Each $ {\Dlt} $ and $ {\Dlt}' $ contains $ n $ triplets of indices,   $(\unl{k}_{1}\unl{r}_{1}\unl{m}_{1})$, $(\unl{k}_{2}\unl{r}_{2}\unl{m}_{2}),\ldots$, and $(\unl{k}_{n}\unl{r}_{n}\unl{m}_{n})$: If any triplet contains one and only one of the indices different from zero, then the corresponding value of $ {\Dlt} $ and $ {\Dlt}' $ vanishes. Finally, the dimensionless coupling constant $ g\equiv  g_{m}/\sqrt{R_{1}\cdots R_{n}} $ explicitly emerges in Eqs.~\eqref{eq:FofA-fourier} and \eqref{FKKmodes}, and it is also encountered in the definitions
\begin{align}
 \mc{D}^{(0,\ldots,0)ab}_{\mu}\equiv &\  \dlt^{ab}\partial_{\mu}-gf^{abc}A_{\mu}^{(0,\ldots,0)c}\ , \label{eq:D00}\\
 F^{(0,\ldots,0)a}_{\mu\nu}\equiv &\ \partial_{\mu}\iA{(0,\ldots, 0)a}{\nu}-\partial_{\nu}\iA{(0,\ldots, 0)a}{\mu}\nonumber\\
 & \ +gf^{abc}\iA{(0,\ldots, 0)b}{\mu}\iA{(0,\ldots, 0)c}{\nu}\ .\label{eq:F00}
\end{align}
To clarify the notation, explicit examples, such as $ m=5$, $m=6 $ and $m=7 $ will be treated in the following section.

The gauge structure of the effective field theory Eq.~\eqref{eq:L4YM} is non-trivial, but it can be unravelled by using the point transformation Eq.~\eqref{AM-expand} and the gauge symmetry Eq.~\eqref{gt-n} present in the higher-dimensional theory.  By introducing the Fourier expansion of the different connection components Eq.~\eqref{AM-expand} and gauge parameters Eq.~\eqref{Fsgp} into the gauge transformation of the higher-dimensional theory Eq.~\eqref{gt-n}, and by making use of the orthogonality of the trigonometric functions, one obtains that the transformation for each zero mode is
\begin{align}\label{dltA-0}
\delta\iA{(0,\ldots, 0)a}{\mu}  =&\  \mc{D}^{(0,\ldots,0)ab}_{\mu}\alpha ^{(0,\ldots,0)b}\nonumber\\
& +gf_{abc}{\sum_{\unl{k}_{1}\ldots \unl{k}_{n}}\!}'\iA{(\unl{k}_{1},\ldots,\unl{k}_{n})b}{\mu}\al^{(\unl{k}_{1},\ldots,\unl{k}_{n})c}\ , 
\end{align}
and for all KK excited modes are
\begin{widetext}
\begin{subequations}\label{gtmodes}
\begin{align}
 \delta\iA{(\unl{m}_{1},\ldots, \unl{m}_{n})a}{\mu} =& \  gf_{abc} \iA{(\unl{m}_{1},\ldots, \unl{m}_{n})b}{\mu} \alpha ^{(0,\ldots,0)c}+\mc{D}^{(0,\ldots,0)ab}_{\mu} \alpha^{(\unl{m}_{1},\ldots, \unl{m}_{n})b}\nonumber\\
 & -gf_{abc}{\sum_{\unl{r}_{1}\ldots \unl{r}_{n}}\!}'{\sum_{\unl{k}_{1}\ldots \unl{k}_{n}}\!\!}'\iA{(\unl{k}_{1},\ldots,\unl{k}_{n})c}{\mu}\al^{(\unl{r}_{1},\ldots,\unl{r}_{n})b}{\Dlt}_{\unl{m}_{1}\ldots \unl{m}_{n}\unl{k}_{1}\ldots \unl{k}_{n}\unl{r}_{1},\ldots \unl{r}_{n}}\ ,\label{dltA-fouriermu} \\
\delta\iA{(\unl{m}_{1},\ldots, \unl{m}_{n})a}{\bar{\mu}} =&\ gf_{abc}\iA{(\unl{m}_{1},\ldots, \unl{m}_{n})b}{\bar{\mu}} \alpha ^{(0,\ldots,0)c} -2\pi\left(\frac{\unl{m}_{1}\dlt_{\bar{\mu}\,5}}{R_{1}}+\cdots+\frac{\unl{m}_{n}\dlt_{\bar{\mu}\,n+4}}{R_{n}} \right) \al^{(\unl{m}_{1},\ldots, \unl{m}_{n})a}\ ,\nonumber\\
& -gf_{abc}{\sum_{\unl{r}_{1}\ldots \unl{r}_{n}}\!}'{\sum_{\unl{k}_{1}\ldots \unl{k}_{n}}\!\!}'\iA{(\unl{k}_{1},\ldots,\unl{k}_{n})c}{\bar{\mu}}\al^{(\unl{r}_{1},\ldots,\unl{r}_{n})b}{\Dlt}'_{\unl{m}_{1},\ldots \unl{m}_{n}\unl{k}_{1}\ldots \unl{k}_{n}\unl{r}_{1}\ldots \unl{r}_{n}}\ . \label{dltA-fourierbarmu}
\end{align}
\end{subequations}
Due to our notation, one can find, for example, the explicit form for $ \dlt\iA{(0,m_{2},0,m_{4},0\ldots, 0)a}{{\mu}} $ (with $ m_{2},m_{4}>0 $) from Eq.\reff{dltA-fouriermu}.

Defining
\begin{subequations}\label{cov-der}
\begin{align}
\mc{D}^{(\unl{m}_{1},\ldots, \unl{m}_{n}\unl{r}_{1},\ldots, \unl{r}_{n})ab}_{\mu} =& \ \mc{D}^{(0,\ldots, 0)ab}_{\mu}\,\delta^{\unl{m}_{1}\unl{r}_{1}}
\cdots\delta^{\unl{m}_{n}\unl{r}_{n}} -g f_{abc}{\sum_{\unl{k}_{1}\ldots \unl{k}_{n}}{\!\!}}'\iA{(\unl{k}_{1},\ldots, \unl{k}_{n})c}{\mu}{\Dlt}_{\unl{m}_{1}\ldots \unl{m}_{n}\unl{k}_{1}\ldots \unl{k}_{n}\unl{r}_{1}\ldots \unl{r}_{n}} \label{Dmn-mu}\ ,\\
\mc{D}_{\bar{\mu}}^{(\unl{m}_{1},\ldots, \unl{m}_{n}\unl{r}_{1},\ldots, \unl{r}_{n})ab} =& -2\pi\left(\frac{\unl{m}_{1}\dlt_{\bar{\mu}\,5}}{R_{1}}+\cdots+\frac{\unl{m}_{n}\dlt_{\bar{\mu}\,n+4}}{R_{n}} \right)\,\delta^{ \unl{m}_{1}\unl{r}_{1}}\cdots\delta^{\unl{m}_{n}\unl{r}_{n}}\dlt^{ab}\nonumber\\
& -gf_{abc}{\sum_{\unl{k}_{1}\ldots \unl{k}_{n}}}'\iA{(\unl{k}_{1},\ldots, \unl{k}_{n})c}{\bar{\mu}} {\Dlt}'_{\unl{m}_{1}\ldots \unl{m}_{n}\unl{k}_{1}\ldots \unl{k}_{n}\unl{r}_{1}\ldots \unl{r}_{n}}\ , \label{Dmn-barmu}
\end{align}
\end{subequations}
one can easily see that Eqs.~\eqref{dltA-fouriermu} and~\eqref{dltA-fourierbarmu} take the following suggestive form:
\begin{subequations}\label{A-GTs}
\begin{align}
\delta\iA{(\unl{m}_{1},\ldots, \unl{m}_{n})a}{\mu} &=  gf_{abc} \iA{(\unl{m}_{1},\ldots, \unl{m}_{n})b}{\mu} \alpha ^{(0,\ldots,0)c}+{\sum_{\unl{r}_{1}\ldots \unl{r}_{n}}\!}'\mc{D}^{(\unl{m}_{1},\ldots, \unl{m}_{n}\unl{r}_{1},\ldots, \unl{r}_{n})ab}_{\mu} \alpha^{(\unl{r}_{1},\ldots, \unl{r}_{n})b} \ , \\
\delta\iA{(\unl{m}_{1},\ldots, \unl{m}_{n})a}{\bar{\mu}} &= gf_{abc}\iA{(\unl{m}_{1},\ldots, \unl{m}_{n})b}{\bar{\mu}} \alpha ^{(0,\ldots,0)c}
+{\sum_{\unl{r}_{1}\ldots \unl{r}_{n}}\!}'\mc{D}^{(\unl{m}_{1},\ldots, \unl{m}_{n}\unl{r}_{1},\ldots, \unl{r}_{n})ab}_{\bar{\mu}} \alpha^{(\unl{r}_{1},\ldots ,\unl{r}_{n})b} \ .
\end{align}
\end{subequations}
Eqs.~\eqref{dltA-0}  and~\eqref{A-GTs} consist of two parts: The SGTs,
\begin{subequations}\label{AnSGT}
\begin{align}
\delta_{\!\!\trm{s}}\iA{(0,\ldots, 0)a}{\mu} &= \mc{D}^{(0,\ldots,0)ab}_{\mu}\alpha ^{(0,\ldots,0)b} \ ,\\
\delta_{\!\!\trm{s}}\iA{(\unl{m}_{1},\ldots, \unl{m}_{n})a}{\mu} &=\  gf_{abc} \iA{(\unl{m}_{1},\ldots, \unl{m}_{n})b}{\mu} \alpha ^{(0,\ldots,0)c} \ ,\\
\delta_{\!\!\trm{s}}\iA{(\unl{m}_{1},\ldots, \unl{m}_{n})a}{\bar{\mu}} &= gf_{abc}\iA{(\unl{m}_{1},\ldots, \unl{m}_{n})b}{\bar{\mu}} \alpha ^{(0,\ldots,0)c}\ ,
\end{align}
\end{subequations}
and the NSGTs
\begin{subequations}\label{AnNSGT}
\begin{align}
\delta_{\!\!\trm{ns}}\iA{(0,\ldots, 0)a}{\mu} &=gf_{abc}{\sum_{\unl{k}_{1}\ldots \unl{k}_{n}}\!}'\iA{(\unl{k}_{1},\ldots,\unl{k}_{n})b}{\mu}\al^{(\unl{k}_{1},\ldots,\unl{k}_{n})c}\ , \\
\delta_{\!\!\trm{ns}}\iA{(\unl{m}_{1},\ldots, \unl{m}_{n})a}{\mu} &={\sum_{\unl{k}_{1}\ldots \unl{k}_{n}}\!}'\mc{D}^{(\unl{m}_{1},\ldots, \unl{m}_{n}\unl{k}_{1},\ldots, \unl{k}_{n})ab}_{\mu} \alpha ^{(\unl{k}_{1},\ldots, \unl{k}_{n})b} \ , \\
\delta_{\!\!\trm{ns}}\iA{(\unl{m}_{1},\ldots, \unl{m}_{n})a}{\bar{\mu}} &={\sum_{\unl{k}_{1}\ldots \unl{k}_{n}}\!}'\mc{D}^{(\unl{m}_{1},\ldots, \unl{m}_{n}\unl{k}_{1},\ldots, \unl{k}_{n})ab}_{\bar{\mu}} \alpha ^{(\unl{k}_{1},\ldots ,\unl{k}_{n})b} \ \label{AnNSGT-barmu}.
\end{align}
\end{subequations}
From Eq.~\eqref{AnSGT} the zero mode components $ A^{(0,\ldots,0)a}_{\mu} $ transform as gauge fields with respect to the SGTs; these transformations correspond to the group $ SU(N,\mc{M}^{4}) $. All KK excited modes $ A^{(0,m_{i},0,\ldots,0)a}_{M}  $, $ \ldots$ , $ A^{(m_{1},\ldots,m_{n})a}_{M}  $ transform as matter fields in the adjoint representation of this group. 

At the level of the Fourier components of the curvature, the corresponding transformations can be obtained by plugging the Fourier expansions of curvature components [Eq.~\eqref{FMN-expand-n}] and gauge parameters [Eq.~\eqref{Fsgp}] into the transformation rule~\eqref{gtF-n}. Using the orthogonality of the trigonometric functions, one obtains for each zero mode
\begin{subequations}\label{dltF-0}
\begin{align}
\dlt\mc{F}^{(0,\ldots, 0)a}_{\mu\nu} =&\ gf_{abc}(\mc{F}^{(0,\ldots,0)b}_{\mu\nu}\al^{(0,\ldots,0)c} +{\sum_{\unl{k}_{1}\ldots \unl{k}_{n}}\!}'\mc{F}^{(\unl{k}_{1},\ldots, \unl{k}_{n})b}_{\mu\nu}\al^{(\unl{k}_{1},\ldots, \unl{k}_{n})c})\ , \label{dltF-0mn}\\
\dlt\mc{F}^{(0,\ldots, 0)a}_{\bar{\mu}\bar{\nu}}  =&\ gf_{abc}(\mc{F}^{(0,\ldots,0)b}_{\bar{\mu}\bar{\nu}}\al^{(0,\ldots,0)c} +{\sum_{\unl{k}_{1}\ldots \unl{k}_{n}}\!}'\mc{F}^{(\unl{k}_{1},\ldots, \unl{k}_{n})b}_{\bar{\mu}\bar{\nu}}\al^{(\unl{k}_{1},\ldots, \unl{k}_{n})c})\ , \label{dltF-0bmbn}
\end{align}
\end{subequations}
and for the KK excited modes
\begin{subequations}\label{dltF-modes}
\begin{align}
\dlt\mc{F}^{(\unl{m}_{1},\ldots,\unl{m}_{n})a}_{\mu\nu} = & \ gf_{abc}\left(\mc{F}^{(\unl{m}_{1},\ldots,\unl{m}_{n})b}_{\mu\nu}\al^{(0,\ldots, 0)c}+\ph{{\sum_{m_{1}\ldots m_{n}}\!\!}'}\right.\nonumber\\
&\ +\left. {\sum_{\ \unl{k}_{1}\ldots \unl{k}_{n}}\!\!\!}'\Big[\dlt_{\unl{m}_{1}\unl{k}_{1}}\cdots \dlt_{\unl{m}_{n}\unl{k}_{n}}\mc{F}^{(0,\ldots, 0)b}_{\mu\nu}+{\sum_{\unl{r}_{1}\ldots \unl{r}_{n}}\!\!}'{\Dlt}_{\unl{m}_{1}\ldots \unl{m}_{n}\unl{r}_{1}\ldots \unl{r}_{n}\unl{k}_{1}\ldots \unl{k}_{n}}\mc{F}^{(\unl{r}_{1},\ldots, \unl{r}_{n})b}_{\mu\nu}\Big]\,\al^{(\unl{k}_{1},\ldots, \unl{k}_{n})c}\right)\label{dltF-modes-mn}\\
\dlt\mc{F}^{(\unl{m}_{1},\ldots,\unl{m}_{n})a}_{\mu\bar{\nu}} = & \ gf_{abc}\left(\mc{F}^{(\unl{m}_{1},\ldots,\unl{m}_{n})b}_{\mu\bar{\nu}}\al^{(0,\ldots, 0)c}+{\sum_{\unl{k}_{1}\ldots \unl{k}_{n}}\!\!}'{\sum_{\ \unl{r}_{1}\ldots \unl{r}_{n}}\!\!\!}'{\Dlt}'_{\unl{m}_{1}\ldots \unl{m}_{n}\unl{r}_{1}\ldots \unl{r}_{n}\unl{k}_{1}\ldots \unl{k}_{n}}\mc{F}^{(\unl{r}_{1},\ldots, \unl{r}_{n})b}_{\mu\bar{\nu}}\al^{(\unl{k}_{1},\ldots, \unl{k}_{n})c}\right)\label{dltF-modes-barmn}\\
\dlt\mc{F}^{(\unl{m}_{1},\ldots,\unl{m}_{n})a}_{\bar{\mu}\bar{\nu}} = & \ gf_{abc}\left(\mc{F}^{(\unl{m}_{1},\ldots,\unl{m}_{n})b}_{\bar{\mu}\bar{\nu}}\al^{(0,\ldots, 0)c}+\ph{{\sum_{m_{1}\ldots m_{n}}\!\!}'}\right.\nonumber\\
&\ +\left. {\sum_{\ \unl{k}_{1}\ldots \unl{k}_{n}}\!\!\!}'\Big[\dlt_{\unl{m}_{1}\unl{k}_{1}}\cdots \dlt_{\unl{m}_{n}\unl{k}_{n}}\mc{F}^{(0,\ldots, 0)b}_{\bar{\mu}\bar{\nu}}+{\sum_{\unl{r}_{1}\ldots \unl{r}_{n}}\!\!}'{\Dlt}_{\unl{m}_{1}\ldots \unl{m}_{n}\unl{r}_{1}\ldots \unl{r}_{n}\unl{k}_{1}\ldots \unl{k}_{n}}\mc{F}^{(\unl{r}_{1},\ldots, \unl{r}_{n})b}_{\bar{\mu}\bar{\nu}}\Big]\,\al^{(\unl{k}_{1},\ldots, \unl{k}_{n})c}\right)\label{dltF-modes-barmbarn}\ .
\end{align}
\end{subequations}
\end{widetext}
Using these transformation laws, the invariance of the effective Lagrangian Eq.~\eqref{eq:L4YM} can be verified. Notice that Eqs.~\eqref{dltF-modes} are valid for any combination of non-negative integers $ \unl{m}_{1},\ldots,\unl{m}_{n} $ except for the case when all of them vanish. Therefore, from Eqs.~\eqref{dltF-modes} one finds, for example, $ \dlt\mc{F}^{(0,m_{2},0,\ldots,0)a}_{\mu\bar{\nu}} $, and so on. In the expansion of the primed sums, the remarks mentioned in the paragraph after Eq.~\eqref{FKKmodes} are also applicable.

It is important to stress that the invariance of the Lagrangian, Eq.~\eqref{eq:L4YM}, under the transformations defined by Eqs.~\eqref{dltA-0} and~\eqref{A-GTs} is not necessarily immediate. Just as in the case of only one UED (see Ref.~\cite{LMNT}), a direct calculation of the curvature variations $\dlt \mc{F}^{(\,\ldots\,)a}_{MN}$ from Eqs.~\eqref{dltA-0} and~\eqref{A-GTs}  gives rise to terms quadratic in $ g $ in comparison with Eq.~\eqref{dltF-modes}; this would destroy the invariance of the effective Lagrangian Eq.~\eqref{eq:L4YM} under gauge transformations~\eqref{dltA-0} and~\eqref{A-GTs}. However, it can be shown that these terms must vanish by consistency with the Fourier transformation Eq.~\eqref{AM-expand}, or by consistency with the canonical transformation introduced in the following section.
\begin{widetext}
Due to the structure of Eqs.~\eqref{eq:FofA-fourier}-\eqref{FKKmodes}, SGTs~\eqref{AnSGT}, and the compactified Lagrangian~\eqref{eq:L4YM}, the mass spectrum of the basic $ ISO(1,3) $ vector and scalar fields can be arranged as in Tables~\ref{mass-spec-I} ,   \ref{mass-spec-III} and \ref{mass-spec-II}. The tilde ($\ \wt{\,}\ $) on top of various fields will be clarified below.
\begin{table}[H]
\begin{center}
\begin{tabular}[c]{| c | c | c |} 
\hline 
Number of nonzero\ &\  KK Tower \ & \ Squared Mass of \\
 Fourier indices \ &\   \ & each field \\
\hline
0 &  $A^{(0,\ldots,0)a}_{\mu} $ & 0\\
\hline
1 & $ A^{(0,m_{i},0,\ldots,0)a}_{\mu} $ & $ \left(\frac{2\pi m_{i}}{R_{i}}\right)^{2} $ \\
\hline
2 & $ A^{(0,m_{i},0,\ldots,0,m_{j},0)a}_{\mu} $ & $ \left(\frac{2\pi m_{i}}{R_{i}}\right)^{2} + \left(\frac{2\pi m_{j}}{R_{j}}\right)^{2}$ \\
\hline
3 & $ A^{(0,m_{i},0,\ldots,0,m_{j},0,\ldots,0,m_{k},0)a}_{\mu} $ & $ \left(\frac{2\pi m_{i}}{R_{i}}\right)^{2} + \left(\frac{2\pi m_{j}}{R_{j}}\right)^{2}+\left(\frac{2\pi m_{k}}{R_{k}}\right)^{2}$ \\
\hline
 \ & \vdots & \ \\
\hline
$n$ & $ A^{(m_{1},m_{2},\ldots,m_{n})a}_{\mu} $ & $ \left(\frac{2\pi m_{1}}{R_{1}}\right)^{2} + \left(\frac{2\pi m_{2}}{R_{2}}\right)^{2}+\cdots+\left(\frac{2\pi m_{n}}{R_{n}}\right)^{2}$ \\
\hline
\end{tabular}
\end{center}
\caption{Mass spectrum of the different KK towers of vector fields\label{mass-spec-I}}
\end{table}

\begin{table}[H]
\begin{center}
\begin{tabular}[c]{| c | c | c |}
\hline 
Number of nonzero\ &\  KK Tower \ & \ Restriction over the \ \\
 Fourier indices \ &\  \ & \ values of $ \bar{\mu} $\ \\
\hline
$1$ &  $A^{(0,m_{i},0\ldots,0)a}_{\bar{\mu}} $ &  $ \bar{\mu}= i+4 $  \\
\hline
2 & $  \wt{A}^{(0,m_{i},0,\ldots,0,m_{j},0)a}_{\bar{\mu}} $ & $  \bar{\mu}= j+4$ \\
\hline
3 & $ \wt{A}^{(0,m_{i},0,\ldots,0,m_{j},0,\ldots,0,m_{k},0)a}_{\bar{\mu}} $ &  $  \bar{\mu}= k+4$   \\
\hline
 \vdots  &  &  \\
\hline
$n-1$ & $ \wt{A}^{(m_{i_{1}},m_{i_{2}},\ldots,m_{i_{r-1}},0,m_{i_{r+1}},\ldots,m_{i_{n}})a}_{\bar{\mu}} $ & $ \bar{\mu}=i_{n}+4 $\\
\hline
$n$ & $ \wt{A}^{(m_{{1}},m_{{2}},\ldots,m_{{n}})a}_{\bar{\mu}} $ & $ \bar{\mu}=n+4 $ \\
\hline
\end{tabular} 
\end{center}
\caption{Massless scalar fields or pseudo-Goldstone bosons\label{mass-spec-III}}
\end{table}

\begin{table}[H]
\begin{center}
\begin{tabular}[c]{| c | c | c | c |}
\hline 
Number of nonzero\ &\  KK Tower \ & \ Restriction over the \ & \ Squared Mass of \\
 Fourier indices \ &\  \ & \ values of $ \bar{\mu} $\ & \ each field \\
\hline
$1$ &  $A^{(0,m_{i},0\ldots,0)a}_{\bar{\mu}} $ &  $ \bar{\mu}\neq i+4 $ & $\left(\frac{2\pi m_{i}}{R_{i}}\right)^{2}$ \\
\hline
2 & $  \wt{A}^{(0,m_{i},0,\ldots,0,m_{j},0)a}_{\bar{\mu}} $ & $  \bar{\mu}= i+4$ & $\left(\frac{2\pi m_{i}}{R_{i}}\right)^{2}+\left(\frac{2\pi m_{j}}{R_{j}}\right)^{2}$ \\
\hline
2 & $  A^{(0,m_{i},0,\ldots,0,m_{j},0)a}_{\bar{\mu}} $ & $  \bar{\mu}\neq i+4,j+4$ & $\left(\frac{2\pi m_{i}}{R_{i}}\right)^{2}+\left(\frac{2\pi m_{j}}{R_{j}}\right)^{2}$ \\
\hline
3 & $ \wt{A}^{(0,m_{i},0,\ldots,0,m_{j},0,\ldots,0,m_{k},0)a}_{\bar{\mu}} $ &  $  \bar{\mu}= i+4,j+4$  & $ \left(\frac{2\pi m_{i}}{R_{i}}\right)^{2} + \left(\frac{2\pi m_{j}}{R_{j}}\right)^{2}+\left(\frac{2\pi m_{k}}{R_{k}}\right)^{2}$ \\
\hline
3 & $ A^{(0,m_{i},0,\ldots,0,m_{j},0,\ldots,0,m_{k},0)a}_{\bar{\mu}} $ &  $  \bar{\mu}\neq i+4,j+4,k+4$  & $ \left(\frac{2\pi m_{i}}{R_{i}}\right)^{2} + \left(\frac{2\pi m_{j}}{R_{j}}\right)^{2}+\left(\frac{2\pi m_{k}}{R_{k}}\right)^{2}$ \\
\hline
 \vdots  &  & \ & \\
\hline
$n-1$ & $ \wt{A}^{(m_{i_{1}},m_{i_{2}},\ldots,m_{i_{r-1}},0,m_{i_{r+1}},\ldots,m_{i_{n}})a}_{\bar{\mu}} $ & $ \bar{\mu}\neq i_{r}+4, i_{n}+4 $ &$\sum_{j\neq i_{r}} \left(\frac{2\pi m_{j}}{R_{j}}\right)^{2}$ \\
\hline
$n-1$ & $ A^{(m_{i_{1}},m_{i_{2}},\ldots,m_{i_{r-1}},0,m_{i_{r+1}},\ldots,m_{i_{n}})a}_{\bar{\mu}} $ & $ \bar{\mu}=i_{r}+4 $ &$\sum_{j\neq i_{r}} \left(\frac{2\pi m_{j}}{R_{j}}\right)^{2}$ \\
\hline
$n$ & $ \wt{A}^{(m_{{1}},m_{{2}},\ldots,m_{{n}})a}_{\bar{\mu}} $ & $ \bar{\mu}\neq n+4 $ &$\sum_{j} \left(\frac{2\pi m_{j}}{R_{j}}\right)^{2}$ \\
\hline
\end{tabular}
\end{center}
\caption{Massive scalar fields\label{mass-spec-II}}
\end{table}
\end{widetext}
From Table~\ref{mass-spec-I} there are $ 2^{n}-1 $ different KK towers of massive vector fields; one of these towers is, for example,  represented by $ A^{(0,m_{2},m_{3},0,\ldots,0)a}_{\mu} $, with  $ m_{2},m_{3}=1,2,3,\ldots\ $. 

In the compactified theory, there are $ (2^{n}-1)n $ different KK towers of scalars: $ A^{(0,m_{i},0,\ldots,0)a}_{\bar{\mu}} $, $ A^{(0,m_{i},0,\ldots,0,m_{j},0,\ldots,0)a}_{\bar{\mu}} $, $ \ldots $ , and $ A^{(m_{1},m_{2},\ldots,m_{n})a}_{\bar{\mu}} $. Among the $ \binom{n}{1}n $ different KK towers of scalars of the form $ A^{(0,m_{i},0,\ldots,0)a}_{\bar{\mu}} $, $ \binom{n}{1}(n-1) $ KK towers include massive fields of the type $  A^{(0,m_{i},0,\ldots,0)a}_{\bar{\mu}}  $ ($ \bar{\mu}\neq i+4 $), and the remaining $ \binom{n}{1} $ KK towers contain massless fields of the type $  A^{(0,m_{i},0,\ldots,0)a}_{i+4}  $, all of which straightforwardly emerge from the squared terms $ \mc{F}^{(0,m_{i},0,\ldots,0)a}_{\bar{\mu}\bar{\nu}} \mc{F}^{(0,m_{i},0,\ldots,0)\bar{\mu}\bar{\nu}}_{a} $ in the effective Lagrangian, Eq.~\eqref{eq:L4YM}. Among the $ \binom{n}{2}n $ different KK towers of scalars of the form $ A^{(0,m_{i},0,\ldots,0,m_{j},0)a}_{\bar{\mu}} $, $ \binom{n}{2}(n-2) $ include massive fields of the form $  A^{(0,m_{i},0,\ldots,0,m_{j},0)a}_{\bar{\mu}}  $ ($ \bar{\mu}\neq i+4 $ and $ \bar{\mu}\neq j+4 $), and the remaining $ 2\binom{n}{2} $ KK towers emerge mixed in pairs by $ \binom{n}{2} $ real and symmetric $ 2\times 2 $ matrices within the effective Lagrangian. The diagonalization of these matrices gives rise to $ \binom{n}{2} $ KK towers of massive scalar fields which are denoted by $ \wt{A}^{(0,m_{i},0,\ldots,0,m_{j},0)a}_{\bar{\mu}} $ ($ \bar{\mu}=i+4 $), and $ \binom{n}{2} $ KK towers of massless scalar fields which are denoted by $ \wt{A}^{(0,m_{i},0,\ldots,0,m_{j},0)a}_{\bar{\mu}} $ ($ \bar{\mu}=j+4 $). The eigenvalues in the diagonalization process define the squared mass of the corresponding fields. In the same manner, among the $ \binom{n}{3}n $ KK towers of scalars of the form $ A^{(0,m_{i},0,\ldots,0,m_{j},0,\ldots,0,m_{k},0)a}_{\bar{\mu}} $,  $ \binom{n}{3}(n-3) $ of them contain massive scalar fields of the form $  A^{(0,m_{i},0,\ldots,0,m_{j},0,\ldots,0,m_{k},0)a}_{\bar{\mu}}  $ ($ \bar{\mu}\neq i+4 ,  j+4 ,  k+4 $), and the remaining $ 3\binom{n}{3} $ KK towers emerge mixed in triads by $ \binom{n}{3} $ real and symmetric $ 3\times 3 $ matrices within the effective Lagrangian. The diagonalization of these matrices gives rise to $ 2\binom{n}{3} $ KK towers of massive scalars denoted by $ \wt{A}^{(0,m_{i},0,\ldots,0,m_{j},0,\ldots,0,m_{k},0)a}_{\bar{\mu}} $ ($ \bar{\mu}=i+4 , j+4 $), and $ \binom{n}{3} $ KK towers of massless scalars that include scalars denoted by $ \wt{A}^{(0,m_{i},0,\ldots,0,m_{j},0,\ldots,0,m_{k},0)a}_{\bar{\mu}} $ ($ \bar{\mu}=k+4 $). This classification continues and, in the end, the $ \binom{n}{n}n=n $ KK towers of scalars of the form $ A^{(m_{1},m_{2},\ldots,m_{n})a}_{\bar{\mu}} $ emerged mixed in $ n $-tuples  by $\binom{n}{n}=1 $ real and symmetric $ n\times n $ matrix within the effective Lagrangian. The diagonalization of this matrix gives rise to $ (n-1)\binom{n}{n}=n-1 $ KK towers of massive scalars denoted by $ \wt{A}^{(m_{1},m_{2},\ldots,m_{n})a}_{\bar{\mu}} $ ($ \bar{\mu}\neq n+4 $) and one KK tower of massless scalars denoted by $ \wt{A}^{(m_{1},m_{2},\ldots,m_{n})a}_{\bar{\mu}} $ ($ \bar{\mu}= n+4 $).

Having clarified the origin and meaning of the tilde fields, worth noticing is the equality in the number of different KK towers of massive vector fields ($2^{n}-1  $) and that of massless scalar fields ($ \sum_{k=1}^{n}\binom{n}{k} $), just as it occurs in a genuine Higgs mechanism; thus, we refer to the latter as pseudo-Goldstone bosons. In addition, the number of KK towers, $ \al^{(0,k_{i},0,\ldots,0)a} $, $ \al^{(0,k_{i},0,\ldots,0,k_{j},0,\ldots,0)a}$, $ \ldots $ , $\al^{(k_{1},k_{2},\ldots,k_{n})a} $, involved in the NSGTs Eq.~\eqref{AnNSGT-barmu}, coincides with the number of pseudo-Goldstone bosons cited above. This observation suggests that there is a particular NSGT in which the pseudo-Goldstone bosons can be eliminated; such a particular gauge is the unitary gauge.  There are, however, two remarkable differences between the compactification process we are reporting in this paper and a genuine Higgs mechanism. (i) In the former there are no broken generators at the level of the gauge algebra: the gauge group $ SU(N,\mc{M}^{4}) $ of the compactified theory is a subgroup of the gauge group $ SU(N,\mc{M}^{m}) $ of  the higher-dimensional theory where both structures share the same gauge generators. (ii) In contrast with a Lagrangian where the Higgs mechanism has taken effect, the Lagrangian Eq.~\eqref{eq:L4YM} contains an infinite numerable set of pseudo-Goldstone bosons codified in the corresponding KK towers. 

\subsection{Hamiltonian structure}
\label{ss:B}

This subsection deals with the phase-space description of the model described by Eq.~\eqref{eq:L4YM}. This analysis will allow us to recognize the gauge generators of the SGTs and NSGTs, Eqs.~\eqref{AnSGT} and~\eqref{AnNSGT}, as the  image of the generator of Eq.~\eqref{gt-n} under a particular canonical transformation that is partially given by Eq.~\eqref{AM-expand}. This study is done in the spirit of Ref.~\cite{LMNT}. In order to recognize the desired canonical transformation, let us recall the Hamiltonian description of the theory~Eq.\eqref{YML}, in which the conjugate momentum field to $ \mc{A}^{a}_{M} $ is $ \pi^{M}_{a}=\mc{F}^{M0}_{a} $; the canonical analysis yields the following first-class constraints:
\begin{subequations}\label{YMn-cons}
\begin{align}
\phi^{(1)}_{a} & = \pi^{0}_{a}\approx 0\label{prim-YMn-cons}\\
\phi^{(2)}_{a}  & = \mc{D}^{ab}_{I}\pi^{I}_{a}\approx 0\label{sec-YMn-cons}
\end{align}
\end{subequations}
where $ I $ labels all spatial components of $ \mc{M}^{m} $ and $ \approx $ denotes a weak equality as introduced in Refs.~\cite{dirac_etal}. The number of physical degrees of freedom in Dirac's sense is $ (N^{2}-1)(m-2) $ per spatial point of $ \mc{M}^{m} $. The gauge algebra has the structure that reflects the underlying group: the nontrivial Poisson brackets (PBs) between constraints are
\begin{equation}\label{ga-n}
\{\phi^{(2)}_{a}[u],\phi^{(2)}_{b}[v] \}_{SU(N,\mc{M})}=g_{m}f_{abc}\,\phi^{(2)}_{c}[uv]
\end{equation}
where the PB $ \{\cdot\,,\,\cdot\}_{SU(N,\mc{M})}$ is calculated using the canonical pairs $ (\mc{A}^{a}_{M},\pi^{M}_{a}) $ and $ \phi^{(2)}_{a}[u]:=\int d^{3}x\,d^{n}y\,u(\mb{x},y)\phi^{(2)}_{a}(\mb{x},y) $, with $ u $ a well behaved smearing function. The infinitesimal gauge transformations Eq.~\eqref{gt-n} can be recovered using the gauge generator~\cite{cas82}
\begin{equation}\label{gg-n}
G=(\mc{D}^{ab}_{0}\al^{b})\phi^{(1)}_{a}-\al^{a}\phi^{(2)}_{a}
\end{equation}
via the PBs as follows:
\begin{equation}\label{gg-gt-n}
\dlt \mc{A}^{a}_{\mu}=\{\mc{A}^{a}_{\mu}, G\}_{SU(N,\mc{M})}\ .
\end{equation}

The phase space for the model of Eq.~\eqref{eq:L4YM} is endowed with the conjugate pairs  $(  A^{(0,\ldots,0)a}_{\mu}, \pi^{(0,\ldots,0)\mu}_{a}) $, $(  A^{(0,m_{i},\ldots,0)a}_{M}, \pi^{(0,m_{i},\ldots,0)M}_{a} )$, $ (A^{(0,m_{i},0\ldots,0,m_{j},0,\ldots,0)a}_{M},$ $ \pi^{(0,m_{i},0\ldots,0,m_{j},0,\ldots,0)M}_{a} )$, $ \ldots $ , $ (A^{(m_{1},\ldots,m_{n})a}_{M}, \pi^{(m_{1},\ldots,m_{n})M}_{a}) $; where the different conjugate momentum fields are 
\begin{subequations}\label{pis}
\begin{align}
\pi^{(0,\ldots,0)\mu}_{a} & = \mc{F}^{(0,\ldots,0)\mu 0}_{a}  \ ,\label{pi00}\\
\pi^{(\unl{m}_{1},\ldots,\unl{m}_{n})\mu}_{a} & = \mc{F}^{(\unl{m}_{1},\ldots,\unl{m}_{n})\mu 0}_{a}\ ,\label{pi0m}\\
\pi^{(\unl{m}_{1},\ldots,\unl{m}_{n})\bar{\mu}}_{a} & = \mc{F}^{(\unl{m}_{1},\ldots,\unl{m}_{n})\bar{\mu} 0}_{a}\ . \label{pimn}
\end{align}
\end{subequations}
Recall that the Fourier indices $ (\unl{m}_{1},\ldots,\unl{m}_{n}) $ take values for any combination of non-negative integers except for the case $ (0,0,\ldots,0) $. Therefore, according to Eqs.~\eqref{AnSGT}, \eqref{dltF-0} and ~\eqref{dltF-modes}, canonical pairs correspond to well-defined objects with respect to the group $ SU(N,\mc{M}^{4}) $, whose gauge parameters are $ \al^{(0,\ldots,0)a} $ defined on $ \mc{M}^{4} $. Moreover, the Fourier expansion of the different curvature components Eqs.~\eqref{FMN-expand-n} and their relation with conjugate momenta allow us to write the following expansions:
\begin{subequations}\label{pi-expand}
\begin{align}
\pi^{\mu}_{a}(x,y) =&\ \left(1/\prod_{\al=1}^{n}R_{\al}\right)^{1/2}\pi^{(0,\ldots, 0)\mu}_{a}(x) \nonumber\\
&+ \left(2/\prod_{\al=1}^{n}R_{\al}\right)^{1/2}{\sum_{\unl{m}_{1}\ldots \unl{m}_{n}}\!\!\!}'\,\pi^{(\unl{m}_{1},\ldots , \unl{m}_{n})\mu}_{a}(x)\nonumber\\
&\times\cos\left[ 2\pi\left(\frac{\unl{m}_{1}y^{1}}{R_{1}}+\cdots+\dfrac{\unl{m}_{n}y^{n}}{R_{n}}\right)\right]\ ,  \label{pi-expand-mu}\\
\pi^{\bar{\mu}}_{a}(x,y) =&\ \left(2/\prod_{\al=1}^{n} R_{\al}\right)^{1/2} {\sum _{\unl{m}_{1}\ldots \unl{m}_{n}}\!\!\!}' \pi^{(\unl{m}_{1},\ldots , \unl{m}_{n})\bar{\mu}}_{a}(x)\nonumber\\
&\times\sin\left[2\pi\left(\frac{\unl{m}_{1}y^{1}}{R_{1}}+\cdots+\dfrac{\unl{m}_{n}y^{n}}{R_{n}}\right)\right]\ . \label{pi-expand-barmu}
\end{align}
\end{subequations}
These expansions relate all conjugate momenta of the higher-dimensional theory, manifestly invariant under $ SU(N,\mc{M}^{m}) $, with those inherent to the effective theory invariant under $ SU(N,\mc{M}^{4}) $. Eqs.~\eqref{AM-expand} and~\eqref{pi-expand} together define a canonical transformation; the proof can be carried out along the same lines of the Appendix A of Ref.~\cite{LMNT} and we will not repeat it here. This result ensures that the canonical Hamiltonians of each theory --the fundamental and the compactified one-- are related via this transformation. In addition we have: $ \{\cdot\, ,\cdot\}_{SU(N,\mc{M})}=\{\cdot\, ,\cdot \}_{SU(N,\mc{M}^{4})} $, where $ \{\cdot\, ,\cdot \}_{SU(N,\mc{M}^{4})}  $ is calculated with respect to the built-in canonical conjugate pairs of the compactified theory.  In order to continue with the canonical analysis of the effective theory, we establish the following proposition.
\begin{proposition}\label{prop:pcs}
The canonical transformation induced by the Fourier expansions faithfully maps the set of primary constraints~\eqref{prim-YMn-cons} of the $ m $-dimensional pure $ SU(N,\mc{M}^{m}) $ Yang-Mills theory onto the set of primary constraints 
\begin{subequations}\label{prim-YM4-cons}
\begin{align}
\phi^{(1)(0,\ldots,0)}_{a} & =\pi^{(0,\ldots,0)0}_{a}\approx 0\ ,\\
\phi^{(1)(0,m_{i},0,\ldots,0)}_{a} & = \pi^{(0,m_{i},0,\ldots,0)0}_{a}\approx 0\\
& \ \,\vdots \nonumber\\
\phi^{(1)(m_{1},\ldots, m_{n})}_{a} & = \pi^{(m_{1},\ldots, m_{n})0}_{a}\approx 0
\end{align}
\end{subequations}
associated to the compactified Yang-Mills theory.
\end{proposition}
The proof is straightforward from the Eq.~\eqref{pi-expand-mu} and the linear independence of the trigonometric functions (see also Proposition IV.2 of Ref.~\cite{LMNT}). Proposition~\ref{prop:pcs} implies that the primary Hamiltonian --the generator of time evolution in any constrained system-- in the higher dimensional theory is mapped into the corresponding primary Hamiltonian of the effective theory.  Indeed, a primary Hamiltonian function is defined as the canonical Hamiltonian plus a linear combination of primary constraints. This ensures that the effective theory presents at most secondary constraints which, concurrently, are the image of the secondary constraint Eq.~\eqref{sec-YMn-cons} under the canonical transformation built from Eqs.~\eqref{AM-expand} and~\eqref{pi-expand}. These effective secondary constraints are~\cite{Errata}
\begin{subequations}\label{sec-YM4-cons}
\begin{align}
& \phi^{(2)(0,\ldots,0)}_{a}  =\ \mc{D}^{(0,\ldots,0)ab}_{i}\pi^{(0,\ldots,0)i}_{a}\nonumber\\
&\hspace{2cm}-gf_{abc}{\sum_{\unl{k}_{1}\ldots \unl{k}_{n}}\!\!}'A^{(\unl{k}_{1},\ldots ,\unl{k}_{n})c}_{I}\pi^{(\unl{k}_{1},\ldots, \unl{k}_{n})I}_{b}\\
& \phi^{(2)(\unl{m}_{1},\ldots, \unl{m}_{n})}_{a}  =  {\sum_{\unl{k}_{1}\ldots \unl{k}_{n}}\!}'\Big\lbrace\mc{D}^{(\unl{m}_{1},\ldots,\unl{m}_{n},\unl{k}_{1},\ldots,\unl{k}_{n})ab}_{i}\pi^{(\unl{k}_{1},\ldots,\unl{k}_{n})i}_{b}\nonumber\\
&  +\Big[2\pi\left(\frac{\unl{m}_{1}\dlt_{\bar{\mu}\,5}}{R_{1}}+\cdots+\frac{\unl{m}_{n}\dlt_{\bar{\mu}\,n+4}}{R_{n}} \right)\dlt^{\unl{m}_{1}\unl{k}_{1}}\cdots\dlt^{\unl{m}_{n}\unl{k}_{n}}\dlt^{ab}\nonumber\\
& -gf_{abc}{\sum_{\unl{r}_{1}\ldots \unl{r}_{n}}}'\iA{(\unl{r}_{1},\ldots, \unl{r}_{n})c}{\bar{\mu}} {\Dlt}'_{\unl{r}_{1}\ldots \unl{r}_{n}\unl{k}_{1}\ldots \unl{k}_{n}\unl{m}_{1}\ldots \unl{m}_{n}}\Big]\pi^{(\unl{k}_{1},\ldots, \unl{k}_{n})\bar{\mu}}_{b}\Big\rbrace\nonumber\\
&\ -gf_{abc}A^{(\unl{m}_{1},\ldots, \unl{m}_{n})c}_{i}\pi^{(0,\ldots, 0)i}_{b}\label{seccons}
\end{align}
\end{subequations}
where $ i=1,2,3 $. 

\begin{widetext}
An important consequence of the canonical transformation is that the gauge algebra associated to the effective theory is the image, under such a transformation, of the gauge algebra linked to the higher-dimensional theory~\eqref{ga-n}; hence,
\begin{subequations}\label{ga-4}
\begin{align}
\{\phi^{(2)(0,\ldots,0)}_{a}[u],\phi^{(2)(0,\ldots,0)}_{b}[v]\} & =gf_{abc}\phi^{(2)(0,\ldots,0)}_{c}[uv]\label{ga-4-SGT}\\
\{\phi^{(2)(0,\ldots,0)}_{a}[u],\phi^{(2)(\unl{m}_{1},\ldots, \unl{m}_{n})}_{b}[v]\} & =gf_{abc}\phi^{(2)(\unl{m}_{1},\ldots, \unl{m}_{n})}_{c}[uv] \\
\{\phi^{(2)(\unl{m}_{1},\ldots, \unl{m}_{n})}_{a}[u],\phi^{(2)(\unl{r}_{1},\ldots, \unl{r}_{n})}_{b}[v]\} & =gf_{abc}\,(\dlt_{\unl{m}_{1}\unl{r}_{1}}\cdots\dlt_{\unl{m}_{n}\unl{r}_{n}}\phi^{(2)(0,\ldots,0)}_{c}[uv]\nonumber\\
& \ \ +{\sum _{\unl{k}_{1}\ldots \unl{k}_{n}}\!\!}'{\Dlt}_{\unl{m}_{1},\ldots, \unl{m}_{n}\unl{r}_{1},\ldots, \unl{r}_{n}\unl{k}_{1},\ldots, \unl{k}_{n}}\phi^{(2)(\unl{k}_{1},\ldots, \unl{k}_{n})}_{c}[uv])
\end{align}
\end{subequations}
Finally, the gauge generator that reproduces the gauge transformations Eqs.~\eqref{dltA-0} and~\eqref{A-GTs} is the sum of the following generators:
\begin{subequations}\label{ggs4}
\begin{align}
G_{\!\!\trm{s}}= &\ \big(\mc{D}^{(0,\ldots,0)ab}_{0}\al^{(0,\ldots,0)b}\big)\phi^{(1)(0,\ldots,0)}_{a}+gf_{abc}{\sum_{\unl{m}_{1}\ldots \unl{m}_{n}}\!\!\!}'A^{(\unl{m}_{1},\ldots, \unl{m}_{n})b}_{0}\al^{(0,\ldots,0)c}\phi^{(1)(\unl{m}_{1},\ldots ,\unl{m}_{n})}_{a}
 -\al^{(0,\ldots, 0)a}\phi^{(2)(0,\ldots, 0)}_{a}\ ,\label{gg-SGT}\\
G_{\!\!\trm{ns}}= &\ gf_{abc}{\sum_{\unl{m}_{1}\ldots \unl{m}_{n}}\!\!\!}'\Big(A^{(\unl{m}_{1},\ldots, \unl{m}_{n})b}_{0}\al^{(\unl{m}_{1},\ldots, \unl{m}_{n})c}\phi^{(1)(0,\ldots ,0)}_{a}+
{\sum_{\unl{k}_{1}\ldots \unl{k}_{n}}\!}'\mc{D}^{(\unl{m}_{1},\ldots, \unl{m}_{n},\unl{k}_{1},\ldots, \unl{k}_{n})ab}_{0}\al^{(\unl{k}_{1},\ldots, \unl{k}_{n})b}\phi^{(1)(\unl{m}_{1},\ldots, \unl{m}_{n})}_{a}\nonumber\\
&\ -\al^{(\unl{m}_{1},\ldots, \unl{m}_{n})a}\phi^{(2)(\unl{m}_{1},\ldots, \unl{m}_{n})}_{a}\Big)\ .\label{gg-NSGT}
\end{align}
\end{subequations}
The function $ G_{\!\!\trm{s}} $ ($ G_{\!\!\trm{ns}} $) duly reproduces Eqs.~\eqref{AnSGT} (resp.~Eqs.\eqref{AnNSGT}) via the PBs defined on the phase space associated to the effective theory.
\end{widetext}
$\ $

\newpage

We end this section with the remark that the set of SGTs~\eqref{AnSGT} --in contrast with the set of NSGTs-- does exponentiate into a group, namely, into $ SU(N,\mc{M}^{4}) $ as can be seen from the gauge subalgebra~\eqref{ga-4-SGT} and the fact that the gauge generator $ G_{\!\!\trm{s}} $ contains as parameters $ \al^{(0,0,\ldots,0)a}(x) $. These parameters come from reducing the parameters $ \al^{a}(x,y) $, defined all over $ \mc{M}^{m} $, to $ \mc{M}^{4} $  by compactification [see Eq.~\eqref{Fsgp}]. In terms of the concepts introduced in~Ref.\cite{LMNT} the gauge symmetry of the higher dimensional $ SU(N,\mc{M}^{m}) $ is hidden after the implementation of the canonical transformation into the form of the SGTs and NSGTs.


\section{Special cases:  \texorpdfstring{$ m=5,6,7 $}{}}
\label{examples}
In this section, using the general theory presented above, we recover the main results reported in Ref.~\cite{LMNT} obtained in the OUED ($ m=5 $) context. We also present in some detail the cases $ m=6 $ and $ m=7 $. The last two are particularly useful in order to spell out the notation introduced in the previous section.

\subsection{Case \texorpdfstring{$ m=5 $}{}}

In this special case, $ \mu $ runs from 0 to 3, and $ \bar{\mu} $ only takes the value of 5. Having one extra spatial dimension implies that the relevant quantities in the compactified theory contain only one Fourier-mode index. Inside the primed sum introduced above, the zero Fourier mode is not permitted, hence in five dimensions the following reduction applies:
\begin{equation}\label{5sum'}
\psum{\ \ \,\unl{m}_{1} \ \ \,} T^{(\unl{m}_{1})}S^{(\unl{m}_{1})}=\sum_{m=1}^{\infty}T^{(m)}S^{(m)}\ .
\end{equation}
The general compactified Lagrangian Eq.~\eqref{eq:L4YM} is reduced to the one presented in Refs.~\cite{NT1, LMNT}, namely,
\begin{align}\label{LYM4-5}
\mc{L}_{\!\trm{4YM}}=& -\frac{1}{4}\mc{F}^{(0)a}_{\mu\nu}\mc{F}^{(0)\mu\nu}_{a}\nonumber\\
&-\frac{1}{4}\sum_{m=1}\left(\mc{F}^{(m)a}_{\mu\nu}\mc{F}^{(m)\mu\nu}_{a}+2\mc{F}^{(m)a}_{\mu 5}\mc{F}^{(m)\mu 5}_{a}\right)\ ,
\end{align}
which is a sum of quadratic monomials in the different sectors of the curvature, where each term has a definite Fourier mode.

Curvature components can be expressed in terms of the configuration variables (fields on $ \mc{M}^{4} $) $ A^{(0)a}_{\mu} $, $ A^{(m)a}_{\mu} $, and $ A^{(m)a}_{5} $. It is not difficult to see that in five dimensions, Eqs.~\eqref{eq:FofA-fourier2} and~\eqref{eq:FofA-fourier5} become trivial equalities, and that by using Eq.~\eqref{5sum'} the general expression Eq.~\eqref{eq:FofA-fourier1} straightforwardly becomes:
\begin{align}\label{5FofA-fourier1}
\mc{F}^{(0)a}_{\mu\nu}=F^{(0)a}_{\mu\nu}+gf_{abc}\sum_{k=1}A^{(k)b}_{\mu}A^{(k)c}_{\nu}\ ,
\end{align}
in agreement with the corresponding equation reported in Ref.\cite{LMNT}. In five dimensions, Eq.~\eqref{eq:FofA-fourier3} is directly reduced to
\begin{align*}
\mc{F}^{(\unl{m}_{1})a}_{\mu\nu}=&\ 2\mc{D}^{(0)ab}_{[\mu}A^{(\unl{m}_{1})b}_{\nu]}\nonumber\\
&+gf_{abc}\!\!\!\psum{\ \ \ \unl{r}_{1}\ \ \ }\!\!\!\psum{\ \ \ \unl{k}_{1}\ \ \ } A^{(\unl{k}_{1})b}_{\mu}A^{(\unl{r}_{1})c}_{\nu}\Dlt_{\unl{m}_{1}\unl{k}_{1}\unl{r}_{1}}
\end{align*}
with $ \mc{D}^{(0)ab}_{\mu} $ the obvious reduction from Eq.~\eqref{eq:D00} to five dimensions. According to the notation introduced previously, the foregoing equation is only valid for $ \unl{m}_{1}=m_{1} =1,2,\ldots$~; hence, after performing the primed sums,
\begin{align}\label{5FofA-fourier3}
\mc{F}^{({m}_{1})a}_{\mu\nu}=&\ 2\mc{D}^{(0)ab}_{[\mu}A^{({m}_{1})b}_{\nu]}\nonumber\\
&+gf_{abc}\sum_{{r}_{1}}\sum_{{k}_{1}}A^{({k}_{1})b}_{\mu}
A^{({r}_{1})c}_{\nu}\Dlt_{{k}_{1}{r}_{1}{m}_{1}}\ .
\end{align}
In a similar fashion, Eq.~\eqref{eq:FofA-fourier4} is reduced to
\begin{align}\label{5FofA-fourier4}
\mc{F}^{({m}_{1})a}_{\mu 5}=&\ \mc{D}^{(0)ab}_{\mu}A_{5}^{({m}_{1})b}+\frac{2\pi {m}_{1}}{R}A^{({m}_{1})a}_{\mu}\nonumber\\
&+gf_{abc}
\sum_{{r}_{1}}\sum_{{k}_{1}} A^{({k}_{1})b}_{\mu}A^{({r}_{1})c}_{5}\Dlt'_{{m}_{1}{r}_{1}{k}_{1}}\ ,
\end{align}
where no sum is performed in the second term. In this way, the specialization of all relevant equations at the configuration and phase-space levels given in the previous section can be reduced to the corresponding ones reported in Ref.~\cite{LMNT}; see also the remark~\cite{Errata}.

In the model Eq.~\eqref{LYM4-5}, the following classification of fields arise: $ A^{(0)a}_{\mu} $ is a gauge vector field, $ A^{(m_{1})a}_{\mu} $ is a KK tower of massive vector fields, and $ A^{(m_{1})a}_{5} $ arises as a KK tower of massless $ SO(1,3) $-scalar fields. In this case there is no need to diagonalize any bilinear term in the Lagrangian. 

\subsection{Cases \texorpdfstring{$ m=6 $}{} and \texorpdfstring{$ m=7 $}{}}

In the case $ m=6 $, the fundamental theory is defined on a six-dimensional spacetime where $ \mu=0,1,2,3 $ and  $ \bar{\mu}=5,6 $. Having two extra spatial dimensions means that all relevant fields, in the effective theory, contain two Fourier modes. The already introduced primed sum over these Fourier indices explicitly reads
\begin{align}\label{6sum'}
\psum{\ \unl{m}_{1} \unl{m}_{2}\ } T^{(\unl{m}_{1}\unl{m}_{2})}S^{(\unl{m}_{1}\unl{m}_{2})}=&\ \sum_{m_{1}=1}^{\infty}T^{(m_{1}0)}S^{(m_{1} 0)}\nonumber\\
&+\sum_{m_{2}=1}^{\infty}T^{(0m_{2})}S^{(0m_{2})}\nonumber\\
& +\sum_{m_{1},m_{2}=1}^{\infty}T^{(m_{1}m_{2})}S^{(m_{1} m_{2})}\ .
\end{align}
Compactification of two spatial dimensions gives rise to the following Lagrangian, \cf~Eq.~\eqref{eq:L4YM}:
\begin{widetext}
\begin{equation}\label{L4YM-6}
\begin{aligned}
\mc{L}_{\!\trm{4YM}}= & -\frac{1}{4}\Big[ \mc{F}^{(00)a}_{\mu\nu}\mc{F}^{(00)\mu\nu}_{a}+\mc{F}^{(00)a}_{\bar{\mu}\bar{\nu}}\mc{F}^{(00)\bar{\mu}\bar{\nu}}_{a}+
\sum_{m_{1}=1}\left(\mc{F}^{(m_{1}0)a}_{\mu\nu}\mc{F}^{(m_{1}0)\mu\nu}_{a}+2\mc{F}^{(m_{1}0)a}_{\mu \bar{\nu}}\mc{F}^{(m_{1}0)\mu \bar{\nu}}_{a}+\mc{F}^{(m_{1}0)a}_{\bar{\mu}\bar{\nu}}\mc{F}^{(m_{1}0)\bar{\mu}\bar{\nu}}_{a}\right)\\
& +\sum_{m_{2}=1}\left(\mc{F}^{(0m_{2})a}_{\mu\nu}\mc{F}^{(0m_{2})\mu\nu}_{a}+2\mc{F}^{(0m_{2})a}_{\mu \bar{\nu}}\mc{F}^{(0m_{2})\mu \bar{\nu}}_{a}+\mc{F}^{(0m_{2})a}_{\bar{\mu}\bar{\nu}}\mc{F}^{(0m_{2})\bar{\mu}\bar{\nu}}_{a}\right) \\
& +\sum_{m_{1},m_{2}=1}\left(\mc{F}^{(m_{1}m_{2})a}_{\mu\nu}\mc{F}^{(m_{1}m_{2})\mu\nu}_{a}+2\mc{F}^{(m_{1}m_{2})a}_{\mu \bar{\nu}}\mc{F}^{(m_{1}m_{2})\mu \bar{\nu}}_{a}+\mc{F}^{(m_{1}m_{2})a}_{\bar{\mu}\bar{\nu}}\mc{F}^{(m_{1}m_{2})\bar{\mu}\bar{\nu}}_{a}\right)\Big] \ .
\end{aligned}
\end{equation}
It is remarkable that each term has a definite Fourier mode. Notice that this model does not coincide with more involved compactification schemes, such as those based on $ T^{2}/Z_{2} $ or the chiral square~\cite{LN}. In this model, the configuration variables explicitly correspond to $ A^{(00)a}_{\mu} $, $ A^{(m_{1}0)a}_{\mu} $, $ A^{(0m_{2})a}_{\mu} $, $ A^{(m_{1}m_{2})a}_{\mu} $, $ A^{(m_{1}0)a}_{\bar{\mu}} $, $ A^{(0m_{2})a}_{\bar{\mu}} $, and $ A^{(m_{1}m_{2})a}_{\bar{\mu}} $. From Eqs.~\eqref{eq:FofA-fourier} and~\eqref{FKKmodes}, each curvature component can be expressed in terms of these variables; using Eq.~\eqref{6sum'}, Eqs.\eqref{eq:FofA-fourier} explicitly read:
\begin{subequations}\label{6FofA-fourier}
\begin{align}
\mc{F}^{(00)a}_{\mu\nu} & = F^{(00)a}_{\mu\nu}+gf_{abc}\Big(\sum_{k_{1}=1}A^{(k_{1}0)b}_{\mu}A^{(k_{1}0)c}_{\nu}+\sum_{k_{2}=1}
A^{(0k_{2})b}_{\mu}A^{(0k_{2})c}_{\nu}+\sum_{k_{1},k_{2}=1} A^{(k_{1}k_{2})b}_{\mu}A^{(k_{1}k_{2})c}_{\nu}\Big)\ ,\label{6FofA-fourier1}\\
\mc{F}^{(00)a}_{\bar{\mu}\bar{\nu}} & = gf_{abc}\Big(\sum_{k_{1}=1}A^{(k_{1}0)b}_{\bar{\mu}}A^{(k_{1}0)c}_{\bar{\nu}}+\sum_{k_{2}=1}
A^{(0k_{2})b}_{\bar{\mu}}A^{(0k_{2})c}_{\bar{\nu}}+\sum_{k_{1},k_{2}=1} A^{(k_{1}k_{2})b}_{\bar{\mu}}A^{(k_{1}k_{2})c}_{\bar{\nu}}\Big)\ , \label{6FofA-fourier2}
\end{align}
\end{subequations}
Eq.~\eqref{eq:FofA-fourier3} comprises the following three equations:
\begin{subequations}\label{6FKKmodes1}
\begin{align}
\mc{F}^{(m_{1}0)a}_{\mu\nu} = & \ 2\mc{D}^{(00)ab}_{[\mu}A^{(m_{1}0)b}_{\nu]}+gf_{abc}\Big(
A^{(k_{1}0)b}_{\mu}A^{(r_{1}0)c}_{\nu}\Dlt_{k_{1}0r_{1}0m_{1}0}\ +\ 
A^{(0k_{2})b}_{\mu}A^{(r_{1}r_{2})c}_{\nu}\Dlt_{0k_{2}r_{1}r_{2}m_{1}0}\  \nonumber\\
& \ + A^{(k_{1}k_{2})b}_{\mu}A^{(0r_{2})c}_{\nu}\Dlt_{k_{1}k_{2}0r_{2}m_{1}0}\ +\ 
A^{(k_{1}k_{2})b}_{\mu}A^{(r_{1}r_{2})c}_{\nu}\Dlt_{k_{1}k_{2}r_{1}r_{2}m_{1}0}\Big)\label{6FofA-fourier3.1}\\
\mc{F}^{(0m_{2})a}_{\mu\nu} = & \ 2\mc{D}^{(00)ab}_{[\mu}A^{(0m_{2})b}_{\nu]}+gf_{abc}\Big(
A^{(k_{1}0)b}_{\mu}A^{(r_{1}r_{2})c}_{\nu}\Dlt_{k_{1}0r_{1}r_{2}0m_{2}}\ +\  
A^{(0k_{2})b}_{\mu}A^{(0r_{2})c}_{\nu}\Dlt_{0k_{2}0r_{2}0m_{2}}\  \nonumber\\
& \ + A^{(k_{1}k_{2})b}_{\mu}A^{(r_{1}0)c}_{\nu}\Dlt_{k_{1}k_{2}r_{1}00m_{2}}\ +\ 
A^{(k_{1}k_{2})b}_{\mu}A^{(r_{1}r_{2})c}_{\nu}\Dlt_{k_{1}k_{2}r_{1}r_{2}0m_{2}}\Big)\label{6FofA-fourier3.2}\\
\mc{F}^{(m_{1}m_{2})a}_{\mu\nu} = & \ 2\mc{D}^{(00)ab}_{[\mu}A^{(m_{1}m_{2})b}_{\nu]}+gf_{abc}\Big(
A^{(k_{1}0)b}_{\mu}A^{(0r_{2})c}_{\nu}\Dlt_{k_{1}00r_{2}m_{1}m_{2}}\ +\
A^{(k_{1}0)b}_{\mu}A^{(r_{1}r_{2})c}_{\nu}\Dlt_{k_{1}0r_{1}r_{2}m_{1}m_{2}}\  \nonumber\\  
& \ + A^{(0k_{2})b}_{\mu}A^{(r_{1}0)c}_{\nu}\Dlt_{0k_{2}r_{1}0m_{1}m_{2}}\ + \ 
A^{(0k_{2})b}_{\mu}A^{(r_{1}r_{2})c}_{\nu}\Dlt_{0k_{2}r_{1}r_{2}m_{1}m_{2}}\ + \ 
A^{(k_{1}k_{2})b}_{\mu}A^{(r_{1}0)c}_{\nu}\Dlt_{k_{1}k_{2}r_{1}0m_{1}m_{2}}\  \nonumber\\
& \ + A^{(k_{1}k_{2})b}_{\mu}A^{(0r_{2})c}_{\nu}\Dlt_{k_{1}k_{2}0r_{2}m_{1}m_{2}}\ +\ 
A^{(k_{1}k_{2})b}_{\mu}A^{(r_{1}r_{2})c}_{\nu}\Dlt_{k_{1}k_{2}r_{1}r_{2}m_{1}m_{2}}\Big)\ .\label{6FofA-fourier3.3}
\end{align}
\end{subequations}
In these expressions the following values for the multi-index object $ \Dlt $ were used:
\begin{align*}
\Dlt_{k_{1}00r_{2}m_{1}0} & =  \Dlt_{k_{1}0r_{1}r_{2}m_{1}0}= \Dlt_{0k_{2}r_{1}0m_{1}0}= \Dlt_{0k_{2}0r_{2}m_{1}0}= \Dlt_{k_{1}k_{2}r_{1}0m_{1}0}=0\ ,
\\
\Dlt_{k_{1}0r_{1}00m_{2}} & =  \Dlt_{k_{1}00r_{2}0m_{2}}= \Dlt_{0k_{2}r_{1}00m_{2}}= \Dlt_{0k_{2}r_{1}r_{2}0m_{2}}= \Dlt_{k_{1}k_{2}0r_{2}0m_{2}}=0\ ,
\\
\Dlt_{k_{1}0r_{1}0m_{1}m_{2}} & =  \Dlt_{0k_{2}0r_{2}m_{1}m_{2}}= 0\ .
\end{align*}
Eq.~\eqref{eq:FofA-fourier4} comprises the following three equations:
\begin{subequations}\label{6FKKmodes2}
\begin{align}
\mc{F}^{(m_{1}0)a}_{\mu\bar{\nu}} = & \ \mc{D}^{(00)ab}_{\mu}A^{(m_{1}0)b}_{\bar{\nu}} + \frac{2\pi m_{1}\dlt_{\bar{\nu}5}}{R_{1}}
A^{(m_{1}0)a}_{\mu}+gf_{abc}\Big( 
A^{(k_{1}0)b}_{\mu}A^{(r_{1}0)c}_{\bar{\nu}}\Dlt'_{m_{1}0r_{1}0k_{1}0}\ +\ 
A^{(0k_{2})b}_{\mu}A^{(r_{1}r_{2})c}_{\bar{\nu}}\Dlt'_{m_{1}0r_{1}r_{2}0k_{2}} \  \nonumber\\
&\ + A^{(k_{1}k_{2})b}_{\mu}A^{(0r_{2})c}_{\bar{\nu}}\Dlt'_{m_{1}00r_{2}k_{1}k_{2}}\ +\ 
A^{(k_{1}k_{2})b}_{\mu}A^{(r_{1}r_{2})c}_{\bar{\nu}}\Dlt'_{m_{1}0r_{1}r_{2}k_{1}k_{2}}\Big) \ ,\label{6FofA-fourier4.1}\\
\mc{F}^{(0m_{2})a}_{\mu\bar{\nu}} = & \ \mc{D}^{(00)ab}_{\mu}A^{(0m_{2})b}_{\bar{\nu}} + \frac{2\pi m_{2}\dlt_{\bar{\nu}6}}{R_{2}}
A^{(0m_{2})a}_{\mu}+gf_{abc}\Big( 
A^{(k_{1}0)b}_{\mu}A^{(r_{1}r_{2})c}_{\bar{\nu}}\Dlt'_{0m_{2}r_{1}r_{2}k_{1}0}\ +\ 
A^{(0k_{2})b}_{\mu}A^{(0r_{2})c}_{\bar{\nu}}\Dlt'_{0m_{2}0r_{2}0k_{2}} \  \nonumber\\
&\ + A^{(k_{1}k_{2})b}_{\mu}A^{(r_{1}0)c}_{\bar{\nu}}\Dlt'_{0m_{2}r_{1}0k_{1}k_{2}}\ +\ 
A^{(k_{1}k_{2})b}_{\mu}A^{(r_{1}r_{2})c}_{\bar{\nu}}\Dlt'_{0m_{2}r_{1}r_{2}k_{1}k_{2}}\Big)\ ,\label{6FofA-fourier4.2}\\
\mc{F}^{(m_{1}m_{2})a}_{\mu\bar{\nu}} = & \ \mc{D}^{(00)ab}_{\mu}A^{(m_{1}m_{2})b}_{\bar{\nu}} + 2\pi\Big(\frac{m_{1}\dlt_{\bar{\nu}5}}{R_{1}}+\frac{m_{2}\dlt_{\bar{\nu}6}}{R_{2}}\Big)A^{(m_{1}m_{2})a}_{\mu}+gf_{abc}\Big( 
A^{(k_{1}0)b}_{\mu}A^{(0r_{2})c}_{\bar{\nu}}\Dlt'_{m_{1}m_{2}0r_{2}k_{1}0}\  \nonumber\\
& \ + A^{(k_{1}0)b}_{\mu}A^{(r_{1}r_{2})c}_{\bar{\nu}}\Dlt'_{m_{1}m_{2}r_{1}r_{2}k_{1}0}\ +\
A^{(0k_{2})b}_{\mu}A^{(r_{1}0)c}_{\bar{\nu}}\Dlt'_{m_{1}m_{2}r_{1}00k_{2}} \ + \ 
A^{(0k_{2})b}_{\mu}A^{(r_{1}r_{2})c}_{\bar{\nu}}\Dlt'_{m_{1}m_{2}r_{1}r_{2}0k_{2}} \  \nonumber\\
&\ + A^{(k_{1}k_{2})b}_{\mu}A^{(r_{1}0)c}_{\bar{\nu}}\Dlt'_{m_{1}m_{2}r_{1}0k_{1}k_{2}}\ +\ 
A^{(k_{1}k_{2})b}_{\mu}A^{(0r_{2})c}_{\bar{\nu}}\Dlt'_{m_{1}m_{2}0r_{2}k_{1}k_{2}}\ +\ 
A^{(k_{1}k_{2})b}_{\mu}A^{(r_{1}r_{2})c}_{\bar{\nu}}\Dlt'_{m_{1}m_{2}r_{1}r_{2}k_{1}k_{2}}\Big)\ , \label{6FofA-fourier4.3}
\end{align}
\end{subequations}
and Eq.~\eqref{eq:FofA-fourier5} comprises the following three equations:
\begin{subequations}\label{6FKKmodes3}
\begin{align}
\mc{F}^{(m_{1}0)a}_{\bar{\mu}\bar{\nu}} = & \  \frac{4\pi m_{1}\dlt_{5[\bar{\mu}}}{R_{1}} A^{(m_{1}0)a}_{\bar{\nu}]}+gf_{abc}\Big(
A^{(k_{1}0)b}_{\bar{\mu}}A^{(r_{1}0)c}_{\bar{\nu}}\Dlt'_{k_{1}0r_{1}0m_{1}0}\ +\ 
A^{(0k_{2})b}_{\bar{\mu}}A^{(r_{1}r_{2})c}_{\bar{\nu}}\Dlt'_{0k_{2}r_{1}r_{2}m_{1}0}\  \nonumber\\
& \ + A^{(k_{1}k_{2})b}_{\bar{\mu}}A^{(0r_{2})c}_{\bar{\nu}}\Dlt'_{k_{1}k_{2}0r_{2}m_{1}0}\ +\
A^{(k_{1}k_{2})b}_{\bar{\mu}}A^{(r_{1}r_{2})c}_{\bar{\nu}}\Dlt'_{k_{1}k_{2}r_{1}r_{2}m_{1}0}\Big) \ ,\label{6FofA-fourier5.1}\\
\mc{F}^{(0m_{2})a}_{\bar{\mu}\bar{\nu}} = & \ \frac{4\pi m_{2}\dlt_{6[\bar{\mu}}}{R_{2}} A^{(0m_{2})a}_{\bar{\nu}]}+gf_{abc}\Big(
A^{(k_{1}0)b}_{\bar{\mu}}A^{(r_{1}r_{2})c}_{\bar{\nu}}\Dlt'_{k_{1}0r_{1}r_{2}0m_{2}}\ +\ 
A^{(0k_{2})b}_{\bar{\mu}}A^{(0r_{2})c}_{\bar{\nu}}\Dlt'_{0k_{2}0r_{2}0m_{2}}\  \nonumber\\
& \ + A^{(k_{1}k_{2})b}_{\bar{\mu}}A^{(r_{1}0)c}_{\bar{\nu}}\Dlt'_{k_{1}k_{2}r_{1}00m_{2}}\ +\
A^{(k_{1}k_{2})b}_{\bar{\mu}}A^{(r_{1}r_{2})c}_{\bar{\nu}}\Dlt'_{k_{1}k_{2}r_{1}r_{2}0m_{2}}\Big)  \ ,\label{6FofA-fourier5.2}\\
\mc{F}^{(m_{1}m_{2})a}_{\bar{\mu}\bar{\nu}} = & \ 4\pi\Big(\frac{m_{1}\dlt_{5[\bar{\mu}}}{R_{1}}+\frac{m_{2}\dlt_{6[\bar{\mu}}}{R_{2}}\Big) A^{(m_{1}m_{2})a}_{\bar{\nu}]}+gf_{abc}\Big(
A^{(k_{1}0)b}_{\bar{\mu}}A^{(0r_{2})c}_{\bar{\nu}}\Dlt'_{k_{1}00r_{2}m_{1}m_{2}}\ +\ 
A^{(k_{1}0)b}_{\bar{\mu}}A^{(r_{1}r_{2})c}_{\bar{\nu}}\Dlt'_{k_{1}0r_{1}r_{2}m_{1}m_{2}}\  \nonumber\\
&\ + A^{(0k_{2})b}_{\bar{\mu}}A^{(r_{1}0)c}_{\bar{\nu}}\Dlt'_{0k_{2}r_{1}0m_{1}m_{2}}\ +\
A^{(0k_{2})b}_{\bar{\mu}}A^{(r_{1}r_{2})c}_{\bar{\nu}}\Dlt'_{0k_{2}r_{1}r_{2}m_{1}m_{2}}\ +\
A^{(k_{1}k_{2})b}_{\bar{\mu}}A^{(r_{1}0)c}_{\bar{\nu}}\Dlt'_{k_{1}k_{2}r_{1}0m_{1}m_{2}}\  \nonumber\\
& \ + A^{(k_{1}k_{2})b}_{\bar{\mu}}A^{(0r_{2})c}_{\bar{\nu}}\Dlt'_{k_{1}k_{2}0r_{2}m_{1}m_{2}}\ +\
A^{(k_{1}k_{2})b}_{\bar{\mu}}A^{(r_{1}r_{2})c}_{\bar{\nu}}\Dlt'_{k_{1}k_{2}r_{1}r_{2}m_{1}m_{2}}\Big) \ , \label{6FofA-fourier5.3}
\end{align}
\end{subequations}
where two repeated Fourier indices imply a sum starting from 1. In order to obtain Eqs.~\eqref{6FKKmodes2} and~\eqref{6FKKmodes3}, the following values for the multi-index object $ \Dlt' $ were used:
\begin{align*}
\Dlt'_{m_{1}00r_{2}k_{1}0} & =  \Dlt'_{m_{1}0r_{1}r_{2}k_{1}0}= \Dlt'_{m_{1}0r_{1}00k_{2}}= \Dlt'_{m_{1}00r_{2}0k_{2}}= \Dlt'_{m_{1}0r_{1}0k_{1}k_{2}}=0\ ,
\\
\Dlt'_{0m_{2}r_{1}0k_{1}0} & =  \Dlt'_{0m_{2}0r_{2}k_{1}0}= \Dlt'_{0m_{2}r_{1}00k_{2}}= \Dlt'_{0m_{2}r_{1}r_{2}0k_{2}}= \Dlt'_{0m_{2}0r_{2}k_{1}k_{2}}=0\ ,
\\
\Dlt'_{m_{1}m_{2}r_{1}0k_{1}0} & =  \Dlt'_{m_{1}m_{2}0r_{2}0k_{2}}=0\ .
\end{align*}
In this case, the convenience in the definition of the primed sum symbol is obvious. In fact, it is possible to perform a further reduction in some of the Eqs.~\eqref{6FKKmodes1},~\eqref{6FKKmodes2} and~\eqref{6FKKmodes3} when $ \Dlt_{k_{1}0r_{1}0m_{1}0}=\Dlt_{k_{1}r_{1}m_{1}} $ and $ \Dlt'_{k_{1}0r_{1}0m_{1}0}=\Dlt'_{k_{1}r_{1}m_{1}} $ are taken into account. Using the same line of reasoning, the gauge transformations Eqs.~\eqref{dltA-0} and~\eqref{gtmodes}and all relevant equations in phase space given in subsection ~\ref{ss:B} can be explicitly written down in six dimensions. 

A new aspect in dimensions greater than 5 is the mixing of terms that appear in the compactified Lagrangian which need to be decoupled in order to find the mass spectrum of the theory. To clarify the discussion on this (given in general terms above), let us explicitly spell out how this decoupling mechanism works in the case $ m=6 $. The Fourier component $ A^{(00)a}_{\mu} $ is a gauge vector field. Due to the structure of the Lagrangian in Eq.~\eqref{L4YM-6} and Eqs.~\eqref{6FKKmodes2}, it can be readily seen that $ A^{(m_{1}0)a}_{\mu} $,  $ A^{(0m_{2})a}_{\mu} $ and  $ A^{(m_{1}m_{2})a}_{\mu} $ are three towers of massive vector fields,  with squared masses $ \left(\frac{2\pi m_{1}}{R_{1}}\right)^{2} $, $ \left(\frac{2\pi m_{2}}{R_{2}}\right)^{2} $, and $ \left(\frac{2\pi m_{1}}{R_{1}}\right)^{2} + \left(\frac{2\pi m_{2}}{R_{2}}\right)^{2} $, respectively. Regarding the KK modes in the extra dimensions, the following two towers of massless $ SO(1,3) $-scalars immediately emerge: $ A^{(m_{1}0)a}_{5} $ and $ A^{(0m_{2})}_{6} $. Notice that from the structure of Eqs.~\eqref{6FofA-fourier5.1} and~\eqref{6FofA-fourier5.2},  the following two towers of massive $ SO(1,3) $-scalars immediately emerge: $ A^{(m_{1}0)a}_{6} $ and $ A^{(0m_{2})}_{5} $, with squared masses $ \left(\frac{2\pi m_{1}}{R_{1}}\right)^{2} $ and $ \left(\frac{2\pi m_{2}}{R_{2}}\right)^{2} $, respectively. Finally, from Eq.~\eqref{6FofA-fourier5.3} and the quadratic term $ \mc{F}^{(m_{1}m_{2})a}_{\bar{\mu}\bar{\nu}}\mc{F}^{(m_{1}m_{2})\bar{\mu}\bar{\nu}}_{a} $ in the Lagrangian, the components $ A^{(m_{1}m_{2})a}_{5} $ and $ A^{(m_{1}m_{2})a}_{6}  $ are mixed in the following bilinear form:
\begin{equation}\label{bilinear6}
\begin{pmatrix}
 A^{(m_{1}m_{2})a}_{5} &  A^{(m_{1}m_{2})a}_{6}
\end{pmatrix}
\begin{pmatrix}
\left(\frac{2\pi m_{2}}{R_{2}}\right)^{2} & -\frac{4\pi^{2}m_{1}m_{2}}{R_{1}R_{2}}\\
-\frac{4\pi^{2}m_{1}m_{2}}{R_{1}R_{2}} & \left(\frac{2\pi m_{1}}{R_{1}}\right)^{2}
\end{pmatrix}
\begin{pmatrix}
 A^{(m_{1}m_{2})5}_{a} \\ \\
  A^{(m_{1}m_{2})6}_{a}
\end{pmatrix}\ .
\end{equation}
The diagonalization of the $ 2\times 2 $ matrix involved yields $  \wt{A}^{(m_{1}m_{2})a}_{5} $ ($ m_{1},m_{2}>0 $), representing a KK tower of massive scalar fields with mass  $ \left(\frac{2\pi m_{1}}{R_{1}}\right)^{2} + \left(\frac{2\pi m_{2}}{R_{2}}\right)^{2}$, and $  \wt{A}^{(m_{1}m_{2})a}_{6} $ ($ m_{1},m_{2}>0 $), representing a KK towers  massless scalar fields. In other words, the bilinear form Eq.~\eqref{bilinear6} is unitarily equivalent to
\begin{equation}\label{diag-bilinear6}
\begin{pmatrix}
 \wt{A}^{(m_{1}m_{2})a}_{5} &  \wt{A}^{(m_{1}m_{2})a}_{6}
\end{pmatrix}
\begin{pmatrix}
 \left(\frac{2\pi m_{1}}{R_{1}}\right)^{2} + \left(\frac{2\pi m_{2}}{R_{2}}\right)^{2} & 0\\
0 & 0
\end{pmatrix}
\begin{pmatrix}
 \wt{A}^{(m_{1}m_{2})5}_{a} \\ \\
 \wt{A}^{(m_{1}m_{2})6}_{a}
\end{pmatrix}\ .
\end{equation}
The mixing angle $ \tta $ is
\begin{equation}\label{mix-ang}
\tan\tta=\frac{m_{1}R_{2}}{m_{2}R_{1}}\ .
\end{equation}
In the Appendix \ref{App1}, Tables~\ref{mass-spec6.1}, \ref{mass-spec6.2} and \ref{mass-spec6.3}  summarize the mass spectrum of the compactified theory from six dimensions.

 As the number of extra dimensions increases, the way in which the mixing terms emerge in the Lagrangian becomes more involved. For example, in the case $ m=7 $ --that is, $ 3 $ extra spatial dimensions-- the following bilinear forms appear in the Lagrangian:
\begin{align}\label{bilinear7}
& \begin{pmatrix}
 A^{(m_{1}m_{2}0)a}_{5} &  A^{(m_{1}m_{2}0)a}_{6}
\end{pmatrix}
\begin{pmatrix}
\bt^{2} & -\al\bt\\
-\al\bt & \al^{2}
\end{pmatrix}
\begin{pmatrix}
 A^{(m_{1}m_{2}0)5}_{a} \\ \\
  A^{(m_{1}m_{2}0)6}_{a}
\end{pmatrix}  
\ + \
\begin{pmatrix}
 A^{(m_{1}0m_{3})a}_{5} &  A^{(m_{1}0m_{3})a}_{7}
\end{pmatrix}
\begin{pmatrix}
\gm^{2} & -\al\gm\\
-\al\gm& \al^{2}
\end{pmatrix}
\begin{pmatrix}
 A^{(m_{1}0m_{3})5}_{a} \\ \\
  A^{(m_{1}0m_{3})7}_{a}
\end{pmatrix}\nonumber\\
& \ +
\begin{pmatrix}
 A^{(0m_{2}m_{3})a}_{6} &  A^{(0m_{2}m_{3})a}_{7}
\end{pmatrix}
\begin{pmatrix}
\gm^{2} & -\bt\gm\\
-\bt\gm & \bt^{2}
\end{pmatrix}
\begin{pmatrix}
 A^{(0m_{2}m_{3})6}_{a} \\ \\
 A^{(0m_{2}m_{3})7}_{a}
\end{pmatrix}\nonumber\\
& \ +
\begin{pmatrix}
 A^{(m_{1}m_{2}m_{3})a}_{5} &  A^{(m_{1}m_{2}m_{3})a}_{6} & A^{(m_{1}m_{2}m_{3})a}_{7}
\end{pmatrix}
\begin{pmatrix}
\bt^{2}+\gm^{2} & -\al\bt & -\al\gm\\
-\al\bt & \al^{2}+\gm^{2} & -\bt\gm\\
-\al\gm &  -\bt\gm & \al^{2}+\bt^{2}\\
\end{pmatrix}
\begin{pmatrix}
 A^{(m_{1}m_{2}m_{3})5}_{a} \\ \\
 A^{(m_{1}m_{2}m_{3})6}_{a} \\ \\
 A^{(m_{1}m_{2}m_{3})7}_{a}
\end{pmatrix}
\end{align}
where $ \al\equiv  \left(\frac{2\pi m_{1}}{R_{1}}\right) $, $  \bt\equiv  \left(\frac{2\pi m_{2}}{R_{2}}\right)$ and $  \gm\equiv  \left(\frac{2\pi m_{3}}{R_{3}}\right)  $. These mixing terms turn out to be unitarily equivalent to the following:
\begin{align}\label{diag-bilinear7}
& \begin{pmatrix}
 \wt{A}^{(m_{1}m_{2}0)a}_{5} &  \wt{A}^{(m_{1}m_{2}0)a}_{6}
\end{pmatrix}
\begin{pmatrix}
\al^{2}+\bt^{2} & 0 \\
0 & 0
\end{pmatrix}
\begin{pmatrix}
 \wt{A}^{(m_{1}m_{2}0)5}_{a} \\ \\
  \wt{A}^{(m_{1}m_{2}0)6}_{a}
\end{pmatrix}  
\ + \
\begin{pmatrix}
 \wt{A}^{(m_{1}0m_{3})a}_{5} &  \wt{A}^{(m_{1}0m_{3})a}_{7}
\end{pmatrix}
\begin{pmatrix}
\al^{2}+\gm^{2} & 0\\
0 & 0
\end{pmatrix}
\begin{pmatrix}
 \wt{A}^{(m_{1}0m_{3})5}_{a} \\ \\
  \wt{A}^{(m_{1}0m_{3})7}_{a}
\end{pmatrix}\nonumber\\
& \ +
\begin{pmatrix}
 \wt{A}^{(0m_{2}m_{3})a}_{6} &  \wt{A}^{(0m_{2}m_{3})a}_{7}
\end{pmatrix}
\begin{pmatrix}
\bt^{2}+\gm^{2} & 0\\
0 & 0
\end{pmatrix}
\begin{pmatrix}
 \wt{A}^{(0m_{2}m_{3})6}_{a} \\ \\
 \wt{A}^{(0m_{2}m_{3})7}_{a}
\end{pmatrix}\nonumber\\
& \ +
\begin{pmatrix}
 \wt{A}^{(m_{1}m_{2}m_{3})a}_{5} &  \wt{A}^{(m_{1}m_{2}m_{3})a}_{6} & \wt{A}^{(m_{1}m_{2}m_{3})a}_{7}
\end{pmatrix}
\begin{pmatrix}
\al^{2}+\bt^{2}+\gm^{2} & 0 & 0\\
0 & \al^{2}+\bt^{2}+\gm^{2} & 0\\
0 & 0 & 0\\
\end{pmatrix}
\begin{pmatrix}
 \wt{A}^{(m_{1}m_{2}m_{3})5}_{a} \\ \\
 \wt{A}^{(m_{1}m_{2}m_{3})6}_{a} \\ \\
 \wt{A}^{(m_{1}m_{2}m_{3})7}_{a}
\end{pmatrix}\ .
\end{align}
In this basis, one easily recognizes $ \wt{A}^{(m_{1}m_{2}0)a}_{5} $, $ \wt{A}^{(m_{1}0m_{3})a}_{5} $, $ \wt{A}^{(0m_{2}m_{3})a}_{6} $, $ \wt{A}^{(m_{1}m_{2}m_{3})a}_{5} $ and $ \wt{A}^{(m_{1}m_{2}m_{3})a}_{6} $ as five KK towers of massive  $ SO(1,3) $-scalars. The mass spectrum for the case $ m=7 $ is summarized in Tables~\ref{mass-spec7.1}, \ref{mass-spec7.2} and \ref{mass-spec7.3} in the Appendix~\ref{App1}.
\end{widetext}

\section{Summary and conclusions}
\label{conclu} 

In this paper, we introduced a novel compactification scheme to derive from an $ m $ dimensional pure $ SU(N,\mc{M}^{m}) $ Yang-Mills theory a compactified theory that contains a four dimensional pure $ SU(N,\mc{M}^{4}) $ Yang-Mills theory in its KK zero mode. We proposed as the base higher-dimensional spacetime the Riemannian flat manifold with  geometry $ \mc{M}^{m}=\mc{M}^{4}\times\mc{N}^{n} $, where the $ n $ extra spatial dimensions are thought of as coordinates of the product space $ \mc{N}^{n}=S^{1}/Z_{2}\times S^{1}/Z_{2}\times\cdots S^{1}/Z_{2}  $. This assumption, together with periodicity and parity conditions, allowed us to expand all fields defined on the bulk in terms of KK towers. The gauge invariance of the effective theory was studied and gauge transformations were classified into the SGTs and NSGTs. It turned out that the former constitute the group $ SU(N,\mc{M}^{4}) $ and are defined by the zero mode of the gauge parameters that characterize the Lie group $ SU(N,\mc{M}^{m}) $. As a consequence of the compactification, and due to the accessible structure of the extra dimensions, a full analysis of the mass spectrum in the effective theory was performed. In this regard, it was found that a Higgs-type mechanism operates after compactification since the number of KK towers of massive vector fields and that of KK towers of massless (pseudo-Goldstone) scalars match. The zero mode of the connection of the higher-dimensional theory can be recognized as a $ SU(N,\mc{M}^{4}) $ gauge field, and $ (2^{n}-1)(n-1) $ KK towers of  massive scalars emerge. These massive scalar fields do not emerge in the OUED scenario, and in a Higgs mechanism sense, they correspond to physical scalar fields. 

By performing a phase-space analysis, we have shown that there exists a canonical transformation by which the fundamental and effective theory are two entirely equivalent representations on the same classical model.  In this model it is not difficult to see a significant feature, namely, that as all the different radii $ R_{i} $ of the different various extra spatial dimensions tend to zero, all massive modes become very heavy and therefore kinematically inaccessible in low-energy dynamics.

At a quantum level, the impact of compactified universal extra dimensions on standard model observables must be quantified via radiative corrections, as this class of new physics becomes primarily manifest at the one-loop level. One of the most successful quantization schemes for gauge systems is the BRST path integral in its field-antifield formalism as it is a covariant quantization approach close to Feynman diagrams. The implementation of this method requires the classical gauge structure underlying the theory and a gauge fixing scheme. It is at these two points that the present work will be useful. On the one hand, we have made a deep analysis of the gauge structure of the compactified theory. On the other hand, a systematic way of extracting all massless pseudo-Goldstone scalars in the compactified theory was given, and this becomes crucial in order to define a unitary gauge for future calculations connected with phenomenological implications.


\acknowledgments{The authors acknowledge financial support from CONACYT, and E.~M-P, H.~N-S and J.~J.~T also acknowledge SNI (M\' exico).}

\appendix
\begin{widetext}
\section{Mass spectrum for \texorpdfstring{$ m=6 $}{} and \texorpdfstring{$ m=7 $}{} cases}\label{App1}
In this Appendix the mass spectrum tables associated to the compactified theories from $ m=6 $ and $ m=7 $ dimensions are collected. These are
\vspace*{-.1cm}
\begin{table}[H]
\begin{center}
\begin{tabular}[c]{| c | c | c |}
\hline
Number of nonzero \ &\  KK Tower \ & \ Squared mass of \\
Fourier indices \ &\   \ & \ each field \\
\hline
0 &  $A^{(00)a}_{\mu} $ & 0\\
\hline
1 & $A^{(m_{1}0)a}_{\mu}$ & $\left(\frac{2\pi m_{1}}{R_{1}}\right)^{2}$ \\
\hline
1 & $A^{(0m_{2})a}_{\mu}$ & $\left(\frac{2\pi m_{2}}{R_{2}}\right)^{2}$ \\
\hline
2 & $A^{(m_{1}m_{2})a}_{\mu}$ & $\left(\frac{2\pi m_{1}}{R_{1}}\right)^{2}+\left(\frac{2\pi m_{2}}{R_{2}}\right)^{2}$ \\
\hline
\end{tabular}
\end{center}
\vspace*{-.2cm}
\caption{Mass spectrum of the different KK towers of vector fields in the compactified theory, case $ m=6 $\label{mass-spec6.1}}
\end{table}
\vspace*{-.1cm}
\begin{table}[H]
\begin{center}
\begin{tabular}[c]{| c | c | c |}
\hline 
Number of nonzero\ &\  KK Tower \ & \ Restriction over the \ \\
 Fourier indices \ &\  \ & \ values of $ \bar{\mu} $\ \\
\hline
$1$ &  $A^{(m_{1}0)a}_{5}$ &  $ \bar{\mu}= 1+4$  \\
\hline
1 & $ A^{(0m_{2})a}_{6}$ & $  \bar{\mu}= 2+4$ \\
\hline
2 & $\wt{A}^{(m_{1}m_{2})a}_{6} $ &  $  \bar{\mu}= 2+4$   \\
\hline
\end{tabular}
\end{center}
\vspace*{-.2cm}
\caption{Massless scalar fields or pseudo-Goldstone bosons in compactified theory, case $ m=6 $\label{mass-spec6.2}}
\end{table}

\begin{table}[H]
\begin{center}
\begin{tabular}[c]{| c | c | c | c |}
\hline 
Number of nonzero\ &\  KK Tower \ & \ Restriction over the \ & \ Squared Mass of \\
 Fourier indices \ &\  \ & \ values of $ \bar{\mu} $\ & \ each field \\
\hline
$1$ & $A^{(0m_{2})a}_{5}$  &  $ \bar{\mu}\neq 2+4 $ & $\left(\frac{2\pi m_{2}}{R_{2}}\right)^{2}$ \\
\hline
$1$ & $A^{(m_{1}0)a}_{6}$  &  $ \bar{\mu}\neq 1+4 $ & $\left(\frac{2\pi m_{1}}{R_{1}}\right)^{2}$ \\
\hline
2 & $\wt{A}^{(m_{1}m_{2})a}_{5}$ & $  \bar{\mu}\neq 2+4$ & $\left(\frac{2\pi m_{1}}{R_{1}}\right)^{2}+\left(\frac{2\pi m_{2}}{R_{2}}\right)^{2}$ \\
\hline
\end{tabular}
\end{center}
\vspace*{-.2cm}
\caption{Massive scalar fields, case $ m=6 $\label{mass-spec6.3}}
\end{table}
\begin{table}[H]
\begin{center}
\begin{tabular}[c]{| c | c | c |}
\hline
Number of nonzero \ &\  KK Tower \ & \ Squared mass of \\
Fourier indices \ &\   \ & \ each field \\
\hline
0 &  $A^{(000)a}_{\mu} $ & 0\\
\hline
1 &  $A^{(m_{1}00)a}_{\mu}$ & $\left(\frac{2\pi m_{1}}{R_{1}}\right)^{2}$ \\
\hline
1 & $A^{(0m_{2}0)a}_{\mu}$ & $\left(\frac{2\pi m_{2}}{R_{2}}\right)^{2}$ \\
\hline
1 & $A^{(00m_{3})a}_{\mu}$ & $\left(\frac{2\pi m_{3}}{R_{3}}\right)^{2}$ \\
\hline
2 & $A^{(m_{1}m_{2}0)a}_{\mu}$ & $\left(\frac{2\pi m_{1}}{R_{1}}\right)^{2}+\left(\frac{2\pi m_{2}}{R_{2}}\right)^{2}$  \\
\hline
2 & $A^{(m_{1}0m_{3})a}_{\mu}$ & $\left(\frac{2\pi m_{1}}{R_{1}}\right)^{2}+\left(\frac{2\pi m_{3}}{R_{3}}\right)^{2}$ \\
\hline
2 & $A^{(0m_{2}m_{3})a}_{\mu}$ & $\left(\frac{2\pi m_{2}}{R_{2}}\right)^{2}+\left(\frac{2\pi m_{3}}{R_{3}}\right)^{2}$  \\
\hline
3 & $A^{(m_{1}m_{2}m_{3})a}_{\mu}$ & $\left(\frac{2\pi m_{1}}{R_{1}}\right)^{2}+\left(\frac{2\pi m_{2}}{R_{2}}\right)^{2}+\left(\frac{2\pi m_{3}}{R_{3}}\right)^{2}$   \\
\hline
\end{tabular}
\end{center}
\vspace*{-.2cm}
\caption{Mass spectrum of the different KK towers of vector fields in the compactified theory, case $ m=7 $\label{mass-spec7.1}}
\end{table}
\vspace*{-.5cm}
\begin{table}[H]
\begin{center}
\begin{tabular}[c]{| c | c | c |}
\hline 
Number of nonzero\ &\  KK Tower \ & \ Restriction over the \ \\
 Fourier indices \ &\  \ & \ values of $ \bar{\mu} $\ \\
\hline
$1$ &  $A^{(m_{1}00)a}_{5}$ &  $ \bar{\mu}= 1+4$  \\
\hline
1 & $ A^{(0m_{2}0)a}_{6}$ & $  \bar{\mu}= 2+4$ \\
\hline
1 & $ A^{(00m_{3})a}_{6}$ & $  \bar{\mu}= 3+4$ \\
\hline
2 & $\wt{A}^{(m_{1}m_{2}0)a}_{6} $ &  $  \bar{\mu}= 2+4$   \\
\hline
2 & $\wt{A}^{(m_{1}0m_{3})a}_{7} $ &  $  \bar{\mu}= 3+4$   \\
\hline
2 & $\wt{A}^{(0m_{2}m_{3})a}_{7}$ &  $  \bar{\mu}= 3+4$   \\
\hline
3 & $\wt{A}^{(m_{1}m_{2}m_{3})a}_{7}$ &  $  \bar{\mu}= 3+4$   \\
\hline 
\end{tabular}
\end{center}
\vspace*{-.2cm}
\caption{Massless scalar fields or pseudo-Goldstone bosons in compactified theory, case $ m=7 $\label{mass-spec7.2}}
\end{table}

\begin{table}[H]
\begin{center}
\begin{tabular}[c]{| c | c | c | c |}
\hline 
Number of nonzero\ &\  KK Tower \ & \ Restriction over the \ & \ Squared Mass of \\
 Fourier indices \ &\  \ & \ values of $ \bar{\mu} $\ & \ each field \\
\hline
$1$ & $A^{(m_{1}00)a}_{6}$  &  $ \bar{\mu}\neq 1+4 $ &  $\left(\frac{2\pi m_{1}}{R_{1}}\right)^{2}$ \\
\hline
$1$ & $A^{(m_{1}00)a}_{7}$  &  $ \bar{\mu}\neq 1+4 $ & $\left(\frac{2\pi m_{1}}{R_{1}}\right)^{2}$ \\
\hline
$1$ & $A^{(0m_{2}0)a}_{5}$  &  $ \bar{\mu}\neq 2+4 $ & $\left(\frac{2\pi m_{2}}{R_{2}}\right)^{2}$ \\
\hline
$1$ & $A^{(0m_{2}0)a}_{7}$  &  $ \bar{\mu}\neq 2+4 $ & $\left(\frac{2\pi m_{2}}{R_{2}}\right)^{2}$ \\
\hline
$1$ & $A^{(00m_{3})a}_{5}$  &  $ \bar{\mu}\neq 3+4 $ & $\left(\frac{2\pi m_{3}}{R_{3}}\right)^{2}$ \\
\hline
$1$ & $A^{(00m_{3})a}_{6}$  &  $ \bar{\mu}\neq 3+4 $ & $\left(\frac{2\pi m_{3}}{R_{3}}\right)^{2}$ \\
\hline
2 & $\wt{A}^{(m_{1}m_{2}0)a}_{5}$ & $  \bar{\mu}= 1+4$ & $\left(\frac{2\pi m_{1}}{R_{1}}\right)^{2}+\left(\frac{2\pi m_{2}}{R_{2}}\right)^{2}$ \\
\hline
2 & $\wt{A}^{(0m_{2}m_{3})a}_{6}$ & $  \bar{\mu}= 2+4$ & $\left(\frac{2\pi m_{2}}{R_{2}}\right)^{2}+\left(\frac{2\pi m_{3}}{R_{3}}\right)^{2}$ \\
\hline
2 & $\wt{A}^{(m_{1}0m_{3})a}_{5}$ & $  \bar{\mu}= 1+4$ & $\left(\frac{2\pi m_{1}}{R_{1}}\right)^{2}+\left(\frac{2\pi m_{3}}{R_{3}}\right)^{2}$ \\
\hline
2 & ${A}^{(m_{1}m_{2}0)a}_{7}$ & $  \bar{\mu}\neq 1+4,2+4$ & $\left(\frac{2\pi m_{1}}{R_{1}}\right)^{2}+\left(\frac{2\pi m_{2}}{R_{2}}\right)^{2}$ \\
\hline
2 & ${A}^{(m_{1}0m_{3})a}_{6}$ & $  \bar{\mu}\neq 1+4,3+4$ & $\left(\frac{2\pi m_{1}}{R_{1}}\right)^{2}+\left(\frac{2\pi m_{3}}{R_{3}}\right)^{2}$ \\
\hline
2 & ${A}^{(0m_{2}m_{3})a}_{5}$ & $  \bar{\mu}\neq 2+4,3+4$ & $\left(\frac{2\pi m_{2}}{R_{2}}\right)^{2}+\left(\frac{2\pi m_{3}}{R_{3}}\right)^{2}$ \\
\hline
3 & $\wt{A}^{(m_{1}m_{2}m_{3})a}_{5}$ & $  \bar{\mu}\neq 3+4$ & $\left(\frac{2\pi m_{1}}{R_{1}}\right)^{2}+\left(\frac{2\pi m_{2}}{R_{2}}\right)^{2}+\left(\frac{2\pi m_{3}}{R_{3}}\right)^{2}$ \\
\hline
3 & $\wt{A}^{(m_{1}m_{2}m_{3})a}_{6}$ & $  \bar{\mu}\neq 3+4$ & $\left(\frac{2\pi m_{1}}{R_{1}}\right)^{2}+\left(\frac{2\pi m_{2}}{R_{2}}\right)^{2}+\left(\frac{2\pi m_{3}}{R_{3}}\right)^{2}$ \\
\hline
\end{tabular}
\end{center}
\vspace*{-.2cm}
\caption{Massive scalar fields, case $ m=7 $\label{mass-spec7.3}}
\end{table}
\end{widetext}


\begin{thebibliography}{99}
\bibitem{Errata} Notice that the terms enclosed by squared brackets within Eq.~\eqref{seccons} seem to form $ \mc{D}_{\bar{\mu}}^{(\unl{m}_{1},\ldots,\unl{m}_{n}\unl{k_{1}},\ldots,\unl{k}_{n})ab} $; however, they differ by contractions and signs. This in fact is the source of a mistake in Ref.~\cite{LMNT} in its Eq.~(4.38b), such equation must read as $ \phi^{(2)(n)}_{a}=\mc{D}^{(nm)ab}_{i}\pi^{(m)i}_{b}+\frac{2\pi n}{R}\pi^{(n)5}_{a}-gf_{abc}\,A^{(r)c}_{5}\pi^{(k)5}_{b}\Dlt'_{rkn}-gf_{abc}A^{(n)c}_{i}\pi^{(0)i}_{b} $, where repeated indices indicate a sum starting from 1.  This correct equation for $ \phi^{(2)(n)}_{a} $ ensures the correct gauge generator  $ G_{\!\!\trm{ns}} $ in the Eq.~(4.40b) of Ref.~\cite{LMNT}.
\end{thebibliography}
\end{document}